\documentclass[english,aps,prb,reprint, superscriptaddress, amsmath]{revtex4-2}

\usepackage{fancyhdr,graphicx}
\usepackage{longtable}
\usepackage[table,xcdraw]{xcolor}
\usepackage[normalem]{ulem}
\useunder{\uline}{\ul}{}
\usepackage[T1]{fontenc}

\usepackage{times}
\usepackage{color}
\usepackage[colorlinks,bookmarks=false,citecolor=blue,linkcolor=blue,urlcolor=blue]{hyperref}
\input pdfcolor.tex
\usepackage{colortbl,amsthm,amsmath,amssymb} 
\usepackage{graphicx}
\usepackage{epsfig,color}
\usepackage{verbatim}
\usepackage{dcolumn}
\usepackage[normalem]{ulem}
\usepackage{bm}
\usepackage{xcolor}
\usepackage[]{ulem}

\begin{document}
		
\title{ Direct detection of quasiparticle tunneling with a charge-sensitive superconducting sensor coupled to a waveguide}

\author{Kazi Rafsanjani Amin}
\email{kazir@chalmers.se}
\affiliation{
	Department of Microtechnology and Nanoscience, Chalmers University of Technology, 412 96 Gothenburg, Sweden
}%
	
\author{Axel M. Eriksson}%
\affiliation{
	Department of Microtechnology and Nanoscience, Chalmers University of Technology, 412 96 Gothenburg, Sweden
}%
\author{Mikael Kervinen}%
\email{Present address: VTT Technical Research Centre of Finland Ltd. Tietotie 3, Espoo 02150, Finland}
\affiliation{
	Department of Microtechnology and Nanoscience, Chalmers University of Technology, 412 96 Gothenburg, Sweden
}%
\author{Linus Andersson}%
 \affiliation{
	Department of Microtechnology and Nanoscience, Chalmers University of Technology, 412 96 Gothenburg, Sweden
}%
\author{Robert Rehammar}%
 \affiliation{
	Department of Microtechnology and Nanoscience, Chalmers University of Technology, 412 96 Gothenburg, Sweden
}
\author{Simone Gasparinetti}
\email{simoneg@chalmers.se}
\affiliation{
	Department of Microtechnology and Nanoscience, Chalmers University of Technology, 412 96 Gothenburg, Sweden
}

\begin{abstract}

Detecting quasiparticle tunneling events in superconducting circuits provides information about the population and dynamics of non-equilibrium quasiparticles. 
Such events can be detected by monitoring changes in the frequency of an offset-charge-sensitive superconducting qubit. This monitoring has so far been performed by Ramsey interferometry assisted by a readout resonator.
Here, we demonstrate a quasiparticle detector based on a superconducting qubit directly coupled to a waveguide.
We directly measure quasiparticle number parity on the qubit island by probing the coherent scattering of a microwave tone, offering simplicity of operation, fast detection speed, and a large signal-to-noise ratio.
We observe tunneling rates between 0.8 and $7~\rm{s}^{-1}$, depending on the average occupation of the detector qubit, and achieve a temporal resolution below $10~\mu\rm{s}$ without a quantum-limited amplifier.
Our simple and efficient detector lowers the barrier to perform studies of quasiparticle population and dynamics, facilitating progress in fundamental science, quantum information processing, and sensing.
	
\end{abstract}
	
\maketitle

	
\textit{Introduction -- } In superconductors, quasiparticles (QPs)~\cite{Glazman2021} are fundamental excitations emerging from the ground state of the Cooper-pair condensate. The fraction of broken Cooper pairs, $x_{qp}$, in a superconducting island at temperature $T$ is exponentially suppressed by $\Delta/k_BT$, where $\Delta$ is the superconducting energy gap. However, a wide range of experiments have consistently observed much larger values for $x_{qp}$~\cite{Segall2004, Wang2014, Visser2014, Gruenhaupt2018, Mannila2022}, for example, $x_{qp} \simeq 10^{-9} - 10^{-5}$ at $T\leq20$~mK in aluminium,
	while thermal equilibrium predicts $x_{qp}\simeq10^{-50}$. This discrepancy remains a fundamental puzzle.
The generation of QPs can occur through various direct or mediated Cooper-pair breaking mechanisms, including absorption of high-energy photons~\cite{Pan2022}, interaction with phonons~\cite{Henriques2019, Cardani2021, Iaia2022}, and impact of ionizing radiation~\cite{Wilen2021, Martinis2021}. The processes of QP generation, energy-relaxation~\cite{Jalabert2023}, diffusion, and recombination collectively contribute to determining the observed value of $x_{qp}$.

In superconducting quantum devices, nonequilibrium QPs can contribute to energy relaxation~\cite{Gruenhaupt2018, Gruenhaupt2019, Amin2022}, decoherence~\cite{Siddiqi2021, Aumentado2023, Gruenhaupt2019}, temporary interruption in the flow of persistent supercurrents~\cite{Gusenkova2022},  and nonthermal population of quantum states~\cite{Pop2014}. 
Phonon-mediated QP generation can cause correlated errors in distant qubits~\cite{Wilen2021}, undermining the efficiency of error correction protocols.
At the same time, unwanted QP generation limits the ultimate performance of quantum capacitance detectors~\cite{Shaw2009}, microwave kinetic-inductance detectors~\cite{bockstiegel2014}, and superconducting nanowire single photon detectors\cite{EsmaeilZadeh2021}. Understanding the energetics of QP dynamics and suppressing their unwanted generation are thus extremely important from both a fundamental and a practical point of view.

Time-domain measurements of quasiparticle tunneling rates were first performed in superconducting single-electron transistors by monitoring the parity of the superconducting island using rf reflectometry \cite{Naaman2006,Ferguson2006}, with microsecond resolution. More recently, such measurements have been performed using offset-charge-sensitive transmon-type superconducting circuits~\cite{Koch2007}, exploiting the sensitivity of the fundamental frequency of the transmon to changes in the charge parity of the island \cite{Riste2013,Serniak2018,Serniak2019}. The change in frequency was detected by performing a Ramsey sequence on the qubit, followed by dispersive qubit readout. A variation of this scheme was also employed, in which a change in parity imparts a shift to the resonator frequency, and can therefore be directly detected, jointly with the qubit state, in a single-shot measurement \cite{Serniak2019}.
These dispersive detection schemes have enabled a number of studies~\cite{Connolly2023,Diamond2022,Gordon2022,Iaia2022,Liu2022} demonstrating statistics of QP tunneling~\cite{Serniak2018}, impact of high-energy photon on QP generation~\cite{Liu2022}, phonon-induced Cooper-pair breaking~\cite{Iaia2022}, signatures of thermal-equilibrium distribution of QPs~\cite{Connolly2023}, effect of suppressing environmental radiation flux~\cite{Gordon2022}, and superconducting-gap engineering to suppress QP tunneling rates~\cite{Kamenov2023}. However, their implementation requires either calibrated time-domain pulses ( \textit{i.e.} to perform a Ramsey measurement) or a careful alignment of qubit and resonator transition frequencies (for direct dispersive readout). In addition, real-time detection requires the ability to perform high-fidelity single-shot readout, and the detection bandwidth is ultimately limited by the resonator linewidth, which is limited to about 1 MHz if not at the expense of additional complexity in the design (\textit{e.g.} using Purcell filters~\cite{Chen2023}). These complications hinder the widespread diffusion of studies of QP tunneling rates, in spite of their interest and potential benefits.

In this Letter, we couple a moderately charge-sensitive transmon-type superconducting circuit directly to a waveguide, and demonstrate real-time detection of QP tunneling events using a simple, continuous-wave measurement of the coherent scattering of the transmon into the waveguide. 
The direct coupling between transmon and waveguide~\cite{Astafiev2010, Winkel2020a, Aamir2022} results in fast detection speed. Even without a quantum-limited amplifier, we achieve a signal-to-noise ratio (SNR) of 2 for an integration time of ~10~$\mu$s. We observe that the measured QP tunneling rate depends on the strength of the drive used to perform the measurements. We ascribe this dependence to transmon-state-dependent quasiparticle rates, which have been reported recently~\cite{Connolly2023}.  With its simplicity of operation, fast detection speed, and high SNR, the device presented here can be conveniently employed to study the energetics of QPs, to benchmark the efficiency of QP-poisoning mitigation strategies, and to develop high-energy radiation sensors~\cite{Fink2023}.
	
\begin{figure}[!t]
	\begin{center}
		\includegraphics[width=.45\textwidth]{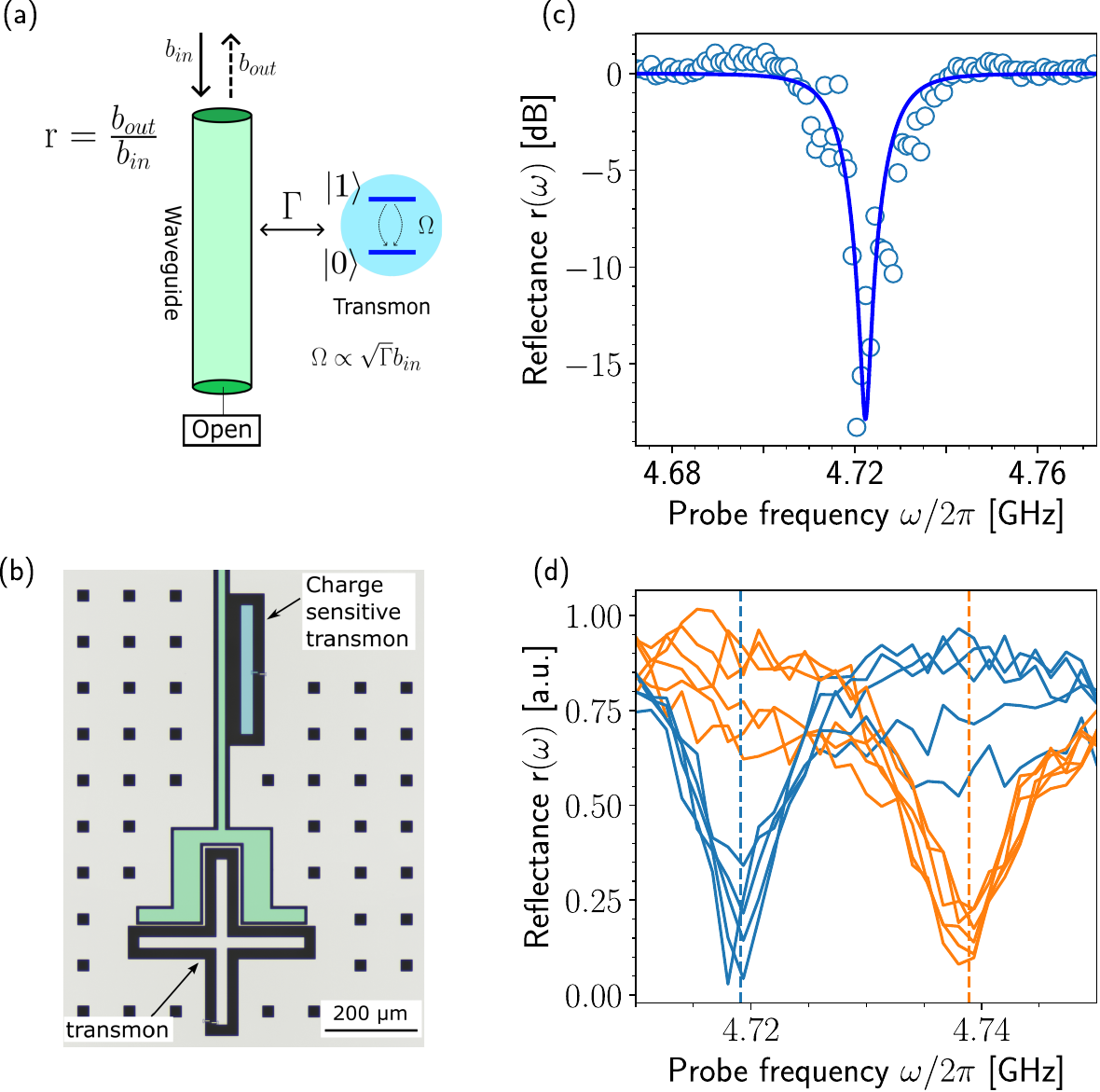}
			\caption{ (a) Schematic of a transmon (approximated as a two-level system) [in blue] coupled to a waveguide [in green], with coupling strength $\Gamma$. One end of the waveguide is terminated with an open, and the device is probed in reflection. 
			(b) A false-colored image of the device where the charge-sensitive transmon is colored in blue, and the waveguide is colored in green. The device includes another transmon, capacitively coupled to the end of the waveguide, which is far detuned in frequency from the charge-sensitive transmon and is not engaged in the present experiment.
			(c) Magnitude of reflection parameter $|r(\omega)|$ measured as a coherent tone is  swept across the $|0\rangle \to|1\rangle$ transition of the transmon, acting as the quasiparticle detector. Destructive interference between reflected and coherently scattered radiation results in a dip at   $|r(\omega_{01})| \to 0$, when the drive power $P_{\rm{in}}$ equates to a Rabi drive rate of $\Omega = 2 \sqrt{\Gamma P_{\rm{in}} /\hbar \omega_{01} } = \Gamma/\sqrt{2}$. The blue solid line is a fit to the data-points (blue open circles). To avoid parity switch because of QP tunneling during acquisition of a single trace, measurements are carried out with smaller number of point in the frequency span and reduced averaging, sufficient enough to resolve the resonance. 
			(d) A total of 11 traces of $|r(\omega)|$, such as in (b), reveals resonances at either of two charge-parity states. The $|0\rangle \to|1\rangle$ transition frequencies for even (odd) parity states at $\omega_{01}^+$ ($\omega_{01}^-$) are marked by vertical orange (blue) dashed lines.
				\label{fig:intro}}	
		\end{center}
	\end{figure}

~\\
\textit{Real-time detection of QP tunneling events --}
In the transmon design, the ratio between Josephson energy, $E_J$ and charging energy, $E_C$ determines the sensitivity of its fundamental transition frequency, $\omega_{01}$, to variations in the offset charge $n_g$, expressed in units of Cooper pairs.  
Tunneling of a QP causes an instantaneous jump of the electron charge parity on the island, $\mathcal{P}=\pm 1$, shifting the offset charge by $1/2$.
For a moderately charge-sensitive transmon, such a jump results in a measurable frequency shift,  depending on the offset charge $n_g$. We deonte the transition frequencies corresponding to parity states $\mathcal{P}=\pm1$ as $\omega_{01}^\pm$.
 Here we design a single-island transmon with $E_J/E_C = $14.5, corresponding to a charge sensitivity $\max_{n_g}|\omega_{01}^+-\omega_{01}^-|/2\pi =20~\rm{MHz}$, and average transition frequency $\bar\omega_{01}/2\pi=4.724~\rm{GHz}$. We capacitively couple the transmon to an open-ended coplanar waveguide (see schematics in Fig.~1(a) and micrograph in Fig.~1(b)) with rate $\Gamma/2\pi=13~\rm{MHz}$.
	
To characterize our device, we send a coherent tone of varying frequency, $\omega_d$, and power, $P_{\rm{in}}$, to the waveguide, driving the fundamental transition of the transmon with a rate $\Omega = 2 \sqrt{\Gamma P_{\rm{in}} /\hbar \omega_{01} }$, and measure the coherently reflected signal to determine the reflection coefficient $r(\omega)$. For simplicity, we perform most of the measurements close to the \textit{magic power}, $P_{\rm m}$, corresponding to a drive rate $\Omega = \Gamma/\sqrt{2}$, at which the reflected field from the open waveguide and the coherently scattered field by the transmon destructively interfere at resonance, giving $|r(0)|\approx 0$.

In Fig.~\ref{fig:intro}(b), we show a representative trace of  $|r(\omega)|$ of our device, measured using a vector network analyzer with a total acquisition time of 1~s. A single resonant dip is observed, to which we fit the theoretical expression for a two-level-system coupled to a waveguide~\cite{Aamir2022, Scigliuzzo2020} to extract the resonant frequency (in this case, $\omega_{01}^-$) and the decay rate, $\Gamma$.

\begin{figure}[!t]
	\begin{center}
		\includegraphics[width=0.485\textwidth]{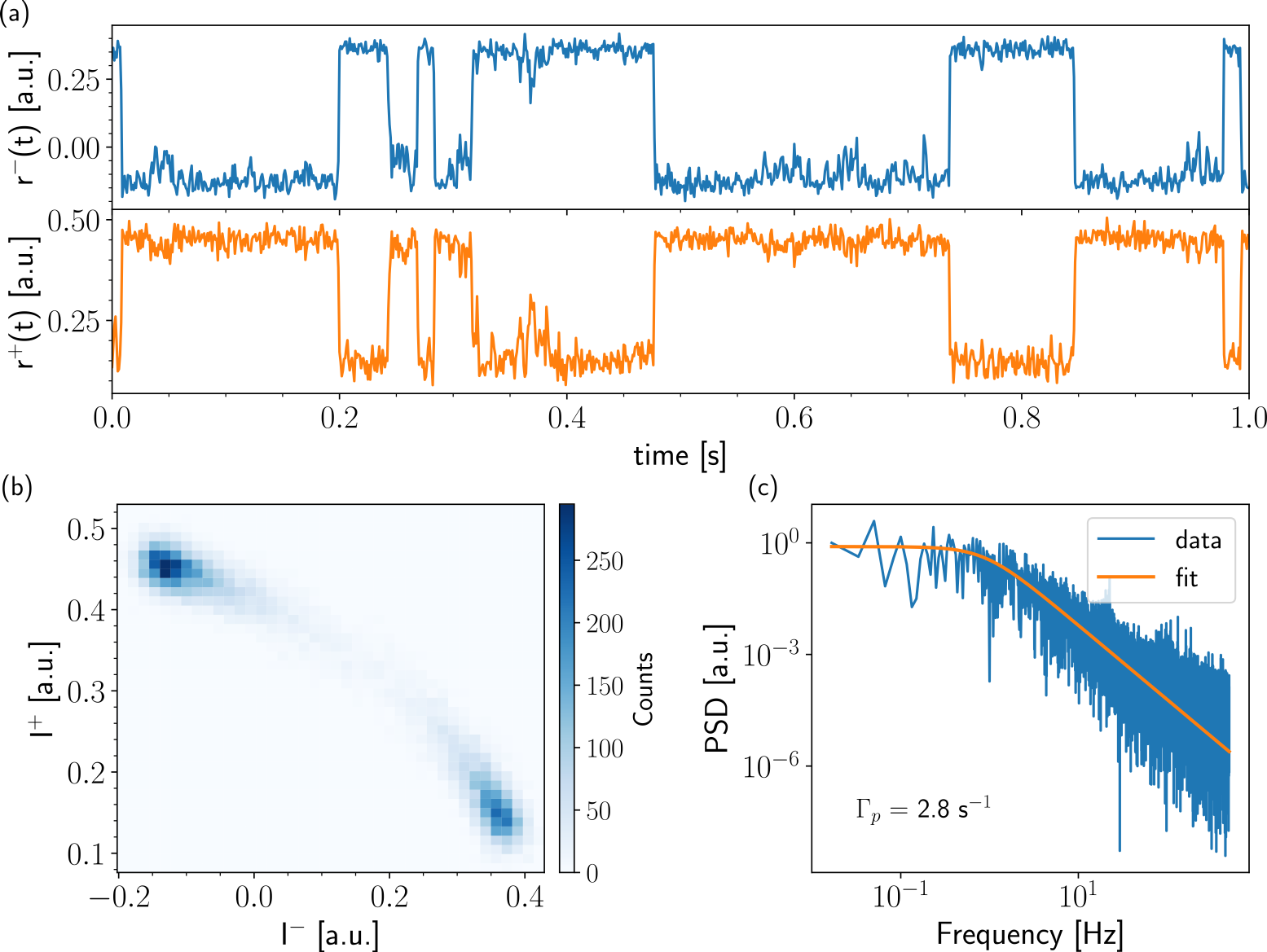}
		\caption{ 
			(a) A 1~s segment time trace of reflection coefficients $r^+$ (top panel) and  $r^-$ (bottom panel), measured while simultaneously driving the even and odd parity states [see Fig.~\ref{fig:intro}(c)].   The traces are anti-correlated random telegraph signal, where a switch between high and low states indicate a QP tunneling event. 
			(b) Two dimensional histogram of 15000 recorded pairs $(r^+,r^-)$, plotted in the I$^+$-I$^-$ plane.
			(c) Power spectral density  (in blue) and fit to the data (orange solid line) using a Lorentzian~\cite{Yuzhelevski2000, Riste2013, Serniak2018}. QP tunneling rate $\Gamma_p$ events are obtained from the corner-frequency of the Lorentzian.
		\label{fig:time_traces}}	
	\end{center}
\end{figure}
	
To detect QP tunneling events, we perform the following sequence of measurements, using a microwave transceiver with both continuous-wave and time-domain capabilities. First, we perform 11 repeated measurements of $|r(\omega)|$ over a frequency span $4\Gamma$, measured with a drive power $P_{\rm m}$. We measure the traces with fast acquisition time ($\sim$100~ms) while optimizing the number of points and the averaging time to achieve large enough SNR to clearly resolve the resonances. As shown in Fig.~\ref{fig:intro}(c), we clearly distinguish two classes of traces (colored by blue and orange), with resonant dips at either of two frequencies. We interpret the stochastic switching between two types of traces as due to QP tunneling events in-between subsequent acquisitions, and identify the two resonant frequencies as $\omega_{01}^\pm$.
 To detect the events in real time, we simultaneously apply two continuous drives at $\omega_{01}^\pm$, record the reflected field in time, and demodulate it at each frequency to extract time traces of the reflection coefficients $r^\pm(t)$. 
Representative traces are shown in Fig.~\ref{fig:time_traces}(a) over a duration of 1~s and with a time binning of 1~ms. Switching events are clearly resolved and manifest themselves as abrupt, perfectly correlated switching of $r^+$ from high to low (low to high) and $r_-$ from low to high (high to low). We confirm the correlation by constructing a 2-dimensional histogram of recorded pairs $(r^-(t),r^+(t))$ [Fig.~2(b)]. Most of the data are clustered around two blobs at opposite ends of the chart. The distance between the blobs far exceeds their broadening due to noise in the measurement, indicating a good signal-to-noise ratio. At the same time, a small fraction of the data lie on an arch connecting the two blobs. These data are consistent with one or more switching events happening during the signal integration window, resulting in a weighted average of the signals associated with the two states. 
Finally, we take the power spectral density (PSD) of one of the recorded traces (say, $r^+(t)$), continuously recorded over 2~minutes, and observe that it is well described by a Lorentzian.
	  
These measurements are consistent with the direct observation of QP tunneling events, which are Poissonian in nature, resulting in a random telegraphic signal in $r^+(t)$ and $r^-(t)$. Since at a given moment in time, the transmon can only have a single $\mathcal{P}$, the measured pairs are anticorrelated. By simultaneously measuring both $r^+$ and $r^-$, we confirm that the transmon stochastically jumps between the two $\mathcal{P}$ states, and we conclude that these jumps originate from of QP tunneling. 
From a Lorentzian fit to the data in Fig.~2(c), we extract the QP tunneling rate, $\Gamma_p$ \cite{Yuzhelevski2000, Riste2013, Serniak2018}. 
	  
\begin {figure}[!t]
	\begin{center}
		\includegraphics[width=0.485\textwidth]{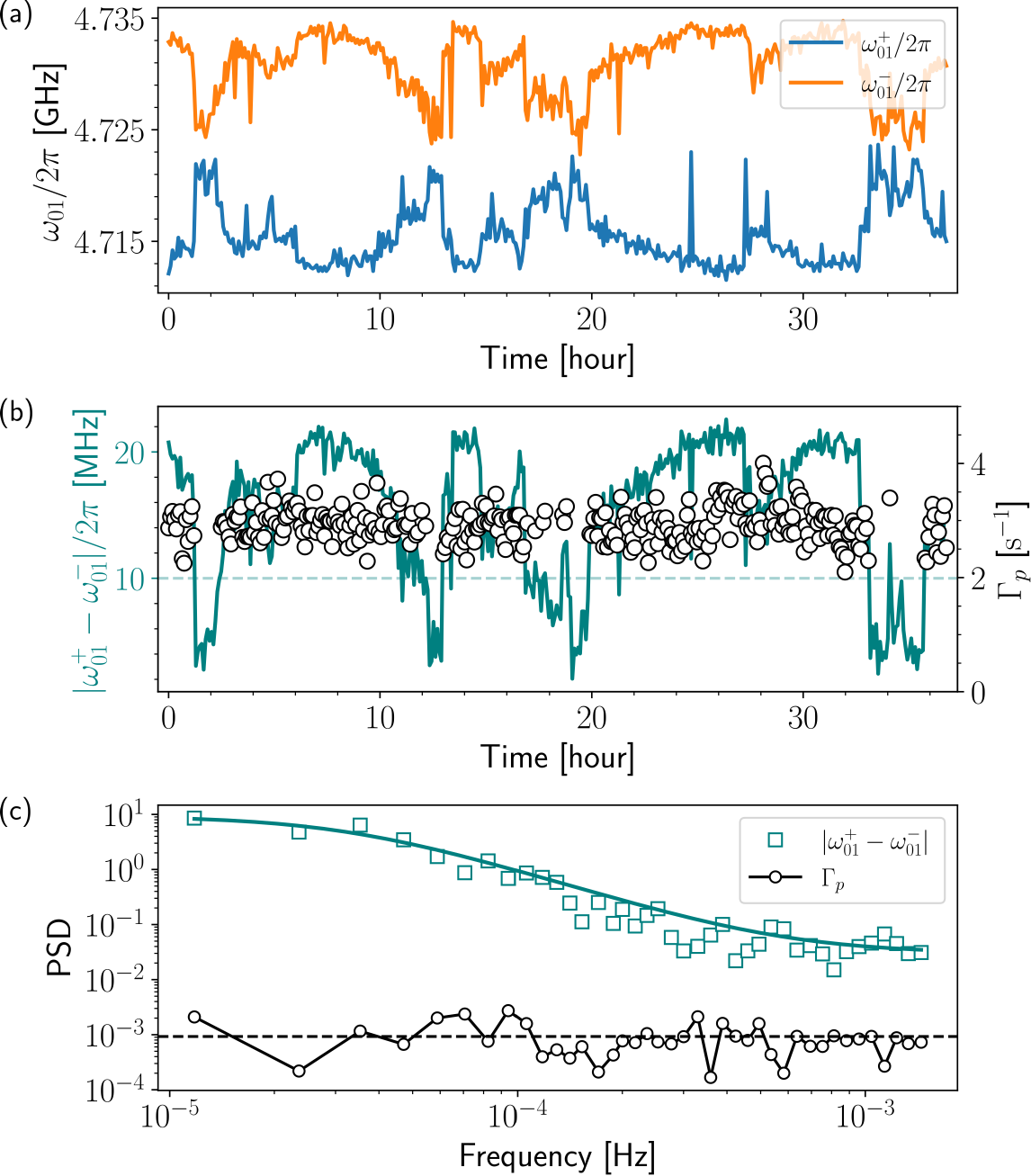}
		\caption{ (a) Time evolution of $|0\rangle \to |1\rangle$  transition frequencies $\omega_{01}^+/2\pi$ (in orange) and $\omega_{01}^-/2\pi$ (in blue) of the two charge-parity states of the QP detector device. Fluctuations in $\omega_{01}^+$ and $\omega_{01}^-$ signifies fluctuations in the background charge landscape in the vicinity of the device. (b) $ |\omega_{01}^+ - \omega_{01}^-|/2\pi$ (left axis,  teal solid line) extracted from the data in (a),  and $\Gamma_p$ (right axis, black circles) versus time.
			(c) Power spectral density of $|\omega_{01}^+ - \omega_{01}^-|/2\pi$ (teal squares) and $\Gamma_p$ (black circles). The teal solid line shows a Lorentzian fit to the power spectral density of $|\omega_{01}^+ - \omega_{01}^-|/2\pi$ versus time, with a corner frequency of $\sim0.2$~mHz.
			\label{fig:stability}}	
	\end{center}
\end {figure}

~\\
\textit{Fluctuations in offset charge and QP tunneling rates --}
We investigate fluctuations by interleaving a measurement of $\omega_{01}^\pm$ [as in Fig.~\ref{fig:intro}(c)] with a 5-min-long record of QP tunneling events [as in Fig.~\ref{fig:time_traces}(a)], from which we compute $\Gamma_p$ [as in Ref.~Fig.~\ref{fig:time_traces}(c)]. By repeating this sequence, we track variations in $\omega_{01}^\pm$ and $\Gamma_p$ over a long period of time [34 hours in Fig.~3(a,b)].

The two resonant frequencies $\omega_{01}^+(t)$ and $\omega_{01}^-(t)$ exhibit slow, perfectly anticorrelated drifts [Fig.~\ref{fig:stability}(a)]. Due to these drifts, the frequency difference $|\omega_{01}^+ - \omega_{01}^-|/2\pi$ [Fig.~3(b), left axis, teal] occasionally takes value smaller than the transmon linewidth $\Gamma$, preventing a clear detection of QP tunneling event. Based on a SNR analysis, we define $|\omega_{01}^+ - \omega_{01}^-|/2\pi\leq10$~MHz as a rejection threshold, below which we do not analyze $r^-(t)$ and $r^+(t)$ traces to extract $\Gamma_p$. 
The measurement record of $\Gamma_p$ shows fluctuations around the mean value of 3 Hz, but appears to be uncorrelated with the large drifts in $|\omega_{01}^+ - \omega_{01}^-|$ [Fig.~\ref{fig:stability}(b), right axis, circles].
We find that the power spectrum of fluctuations of $|\omega_{01}^+ - \omega_{01}^-|$ is Lorentzian, while fluctuations in $\Gamma_p$ are much smaller and exhibit no characteristic time scale [Fig.~\ref{fig:stability}(c), teal open squares and black open circles, respectively].

The frequency gap $|\omega_{01}^+ - \omega_{01}^-|$ can be directly mapped to the offset charge $n_g$, to which we relate the observed drifts~\cite{Pan2022}. These drifts, with characteristic Lorentzian corner frequency of $\sim0.2$~mHz have been observed previously and were ascribed to modifications  in the charge landscape determined by fluctuations between  metastable configurations of one or few two-level systems~\cite{Schloer2019, Tennant2022, Damme2024}. 
We note that it is possible to actively compensate for these drifts by using the measurement of $\omega_{01}^\pm$ to determine the current value of $n_g$, and then readjust $n_g$ to its target value by applying a calibrated dc voltage to a neighboring electrode. This technique is commonly used in single-electron transistors and has been applied also in a context similar to this work~\cite{Connolly2023}.
The lack of correlations between fluctuations in $\Gamma_p$ and $n_g$ suggests that these fluctuations (over the time scales considered) have a completely different origin. In particular, the measured QP tunneling rates are independent of the offset charge and are not affected by slow changes in the electrostatic potential landscape.
	
In addition to the device presented, we have measured two more devices, fabricated on two different wafers, and consistently measured QP tunneling rates $\Gamma_p \sim$2-4~$\rm{s}^{-1}$ (measured at the ``magic power'' $\Omega=\Gamma/\sqrt{2}$) over multiple cooldowns.

\begin {figure}[!t]
	\begin{center}
		\includegraphics[width=0.485\textwidth]{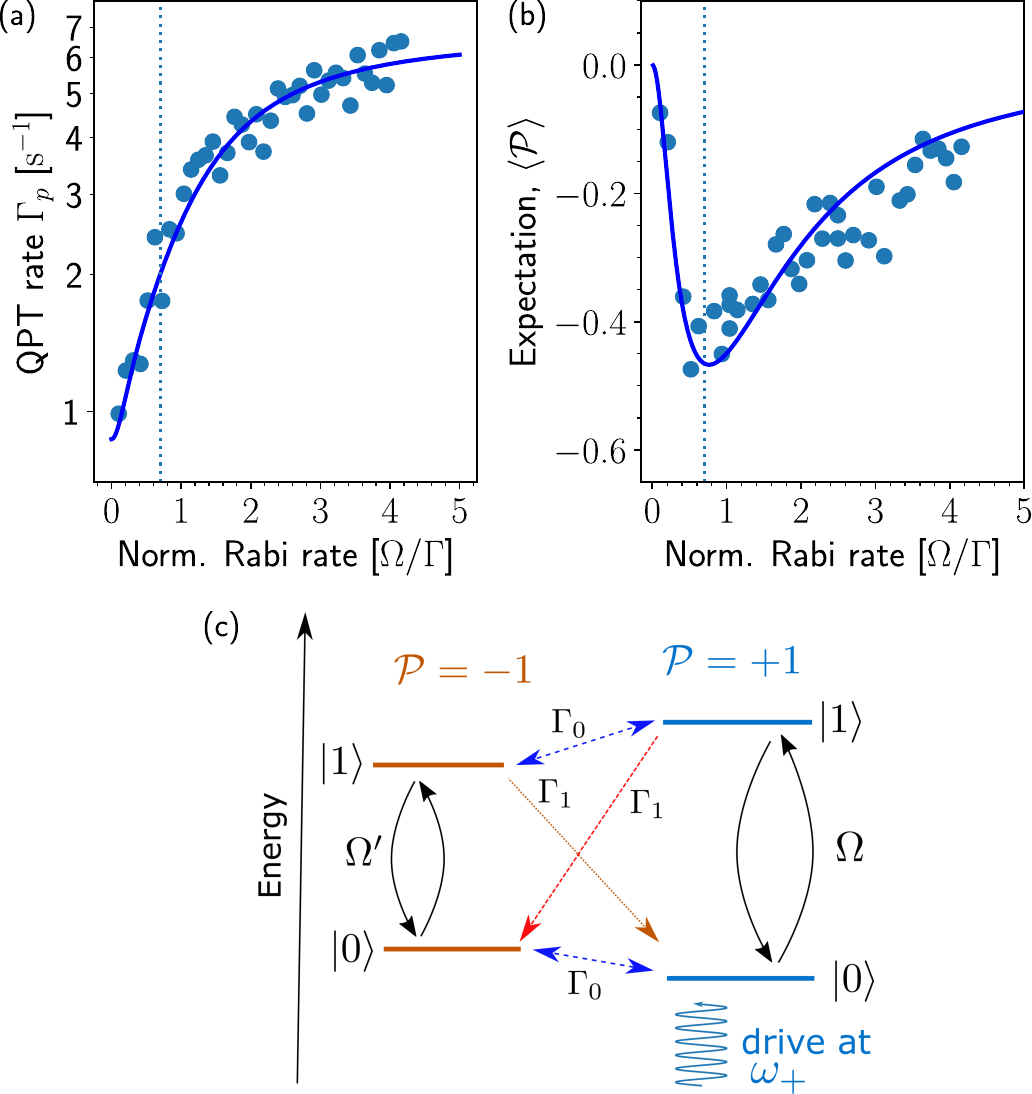}
		\caption{  (a) QP tunneling rate $\Gamma_p$  (blue circles) \textit{vs} amplitude of readout tone, normalized in units of $\Omega/\Gamma$. The vertical dotted line indicates the magic power $\Omega/\Gamma = 1/\sqrt{2}$.  The solid line is a fit to the data points using the model illustrated in (c), with $\Gamma_0$ and $\Gamma_1$ as fit parameters.
		(b) Expectation value of transmon parity, $\langle\mathcal{P}\rangle$ \textit{vs} $\Omega/\Gamma$. The solid line is the corresponding theoretical estimate, using $\Gamma_0$ and $\Gamma_1$ from the fit in (a). 
		(c) Schematic representation of the charge-parity states of the transmon, where the arrows mark allowed transitions: direct transmon drive, with rate $\Omega$, transmon-state independent parity switch, with rate $\Gamma_{0}$, and parity switch assisted by transmon decay, with rate $\Gamma_{1}$. At large drive strengths, due to finite detuning, the non-resonant transition is also effectively driven with a smaller rate, indicated as $\Omega'$.
			\label{fig:snr}}	
	\end{center}
\end {figure}

~\\
\textit{Effect of drive amplitude --}
We further investigate the dependence of the QP tunneling rates on the amplitude of the drive tone used to probe the resonance.  We always maintain $\Omega \ll |\alpha|$, where $\alpha/2\pi\approx -450$~MHz is the anharmonicity of the transmon (design value),  to ensure that  transitions to energy levels  higher than $|1\rangle$  are not driven.
 When driving the $\omega_{01}^+$ transition with a varying rate $\Omega$, we observe significant variations in the measured $\Gamma_p$ [Fig.~\ref{fig:snr}(a)]. $\Gamma_p$ monotonically increases with $\Omega$, with the steepest change observed at low drives ($\Omega/\Gamma \lesssim 1 $) and hints of saturation at large drives ($\Omega/\Gamma \approx 4$). The change in $\Gamma_p$ spans an order of magnitude and the lowest measured rate is $\Gamma_p = 1~\rm{s}^{-1}$. 
We further measure the average parity of the transmon, $\langle\mathcal{P}\rangle$, by using the time traces to determine how much time the transmon spends in each parity state. We find that $\langle\mathcal{P}\rangle$ also depends on $\Omega/\Gamma$ [Fig.~\ref{fig:snr}(b)]. At  low drives, the transmon is equally likely to occupy each of the parity states ($\langle\mathcal{P}\rangle \approx 0$). With increasing drive, $\langle\mathcal{P}\rangle$ first decreases, then drops to a value of -0.45, then slowly increases back to 0. 

We explain both trends as due to the combination of transmon-state-dependent quasiparticle tunneling rates~\cite{Diamond2022, Connolly2023, Krause2024} and varying steady-state qubit population depending on the drive rate [Fig.~\ref{fig:snr}(c)].
In fact, apart from enabling readout of the charge-parity states, the drive tone also induces Rabi oscillations between the ground and the first excited state of the transmon. At low drive strengths, the oscillations are selective to the driven parity state. With increasing drive strength, the steady-state population of the excited state increases until the transition is saturated, and the selectivity decreases.
The observed increase in $\Gamma_p$ with increasing $\Omega$ suggests a larger QP tunneling rate for the excited state of the transmon compared to the ground state. We account for this increase by introducing a tunneling rate $\Gamma_1$, corresponding to a parity switch assisted by  qubit decay~\cite{Connolly2023,Houzet2019,Krause2024}.
The combination of parity-selective drive and qubit state-dependent QP rates preferentially pumps the transmon into the undriven parity state, as observed in Fig.~\ref{fig:snr}(b). When the drive strength far exceeds the frequency splitting between parity states, $\Omega\gg \delta \omega$, the selectivity of the drive is lost and we recover unbiased parity occupation, $\langle\mathcal{P}\rangle=0$.

Based on the model in Fig.~4(c), we calculate the steady-state occupations of the four states [see Supplementary section~S6], from which we determine the average QP tunneling rate $\Gamma_p$ and the average parity $\langle\mathcal{P}\rangle$ (see Supplementary materials). By fitting the model to the data in Fig.~\ref{fig:snr}(a), we reproduce the functional dependence of the data and extract the values $\Gamma_0 = 0.85~\rm{s}^{-1}$ and $\Gamma_1 = 13~\rm{s}^{-1}$. In addition, we find that the same model quantitatively predicts the measured dependence of $\langle\mathcal{P}\rangle$ on $\Omega$ [Fig.~\ref{fig:snr}(b)], without free parameters.

Our observations of state-dependent QP tunneling rates are compatible with Arrhenius-type activated tunneling of quasiparticles trapped in low-superconducting-gap electrodes~\cite{Diamond2022, Connolly2023}. Using the expression $\Gamma_1/\Gamma_0 = \exp(\hbar \omega_q / k_B T)$, we estimate an effective temperature of $\sim83$~mK for the trapped QP bath.

~\\	
\textit{Discussion -- }
We have demonstrated a QP detector based on an offset-charge-sensitive transmon directly coupled to a waveguide. The detection only requires simple, continuous-wave, coherent scattering measurements.  We have measured stable QP tunneling rates between 2 and $4~\rm{s}^{-1}$, uncorrelated to slow fluctuations in the charge offset. By increasing the amplitude of the measurement tone, we observe an increase in the QP tunneling rates, which we attribute to a higher average excited-state population of the transmon. We also observe a biased occupation of the two parity states, which we explain as a combination of parity-state-selective driving and a higher QP tunneling rate when the transmon is excited.
Comparing our theory model to the data, we conclude that the QP tunneling rate of the excited state of the transmon, $\Gamma_1$ is about 15 times larger than transmon-state independent QP tunneling rate, $\Gamma_0$. According to recent literature, tunneling of the QPs trapped in low-superconducting-gap electrodes can be assisted by the energy of the excited state of the transmon, which possibly explains the increase in $\Gamma_1$ as compared to $\Gamma_0$~\cite{Diamond2022, Connolly2023}.

The transmon-state-independent QP tunneling rate, $\Gamma_0$, sets a lower limit to the measured QP tunneling rates. It may result from photon-assisted parity switching~\cite{Houzet2019} or phonon-induced Cooper pair breaking~\cite{Iaia2022}. It is also known to be affected by several factors, including shielding of the package hosting the device~\cite{Saira2012, Gordon2022, Connolly2023}, filtering of the microwave lines used to access it~\cite{Connolly2023}, superconducting gap engineering at the junction electrodes~\cite{VanWoerkom2015,Diamond2022,McEwen2024}, geometry of the transmon pads~\cite{Liu2022}, and presence of normal-metal electrodes acting as phonon traps on the chip~\cite{Iaia2022}.
In Supplementary Table~1, we compare our $\Gamma_p$ with reported values measured with similar devices over the past 6 years, in the range between $10^4$ and $10^{-2}~\rm{s}^{-1}$.
Based on these works, it seems like reducing QP tunneling rate from the $\rm{ms}^{-1}$ to the $\rm{s}^{-1}$ range is a prerequisite to observe transmon-state-dependent QP tunneling rates. In our work, to enter this low-tunneling-rate regime, we protect the device from high-energy photon radiation by using multiple, nested radiation shields at base temperature, and a microwave package designed to minimize radiation leakage~\cite{202Qlab2023} [see Supplementary Fig.~S1]. However, in the presented measurements, we have not employed any indium gaskets (to make the device enclosure light-tight), implementing which in our setup may lead to a further reduction in $\Gamma_p$.

The quantum efficiency for the measurement chain in our setup was $\eta=0.04$  (corresponding to an added noise of 12 photons), limited by the high-mobility electron transistor (HEMT) use as the first amplifier. This efficiency can be improved by using a near-quantum-limited amplification chain, reaching (or even surpassing, in the phase-sensitive mode) $\eta=0.5$~\cite{Walter2017,Lecocq2020,  Ranadive2022}. This improvement would directly translate in an order of magnitude improvement in the temporal resolution, from $10~\mu$s down to below $1~\mu$s. Such high speed and large SNR will be specifically useful for time-resolved detection of correlated parity jumps in distant detectors \cite{McEwen2024,Harrington2024}.

The QP detector transmon presented here can be straightforwardly integrated with other detectors, such as a radiation field thermometer~\cite{Scigliuzzo2020}, measuring in-band thermal photon occupation, and standard transmon qubits coupled to resonators, as probes of various decoherence and dephasing channels~\cite{Burnett2019}. In fact, these devices can all be coupled to the same waveguide [see Supplementary Fig.~S2]. We envision using such a multi-sensor device to benchmark performances of cryogenic components for quantum technologies~\cite{Acharya2023, Pechal2016, Ritter2021}, with the aim to improve the electromagnetic environment used in quantum measurements. 
Furthermore, the direct coupling to the waveguide drastically reduces the footprint of our QP detector compared to conventional readout schemes based on a dispersively coupled resonator. We foresee coupling arrays of these detectors to a common feedline, using frequency multiplexing, and interrogating them with a comb of microwave tones to realize detectors of high-energy radiation~\cite{Fink2023} with 100~$\mu$m spatial and 10~$\mu$s temporal resolution.

~\\
\begin{center}
\textbf{ACKNOWLEDGEMENTS}
\end{center}
~\\
	The authors sincerely acknowledge Gianluigi Catelani and Ioan Pop for valuable discussions.
	The chips were fabricated in the Chalmers Myfab cleanroom.
	K.R.A. is grateful to  Jiaying 	Yang for help with the experimental setup, and to Francois Joint for acquiring micrographs of the device.
	L.A. acknowledge funding from KAW via the WALP (Wallenberg Launch Pad) program.
	S.G. acknowledges financial support from the European Research Council via Grant No.~101041744 ESQuAT.
	This work was supported by the Knut and Alice Wallenberg Foundation via the Wallenberg Centre for Quantum Technology (WACQT).
	
\clearpage
\onecolumngrid

\renewcommand{\theequation}{S\arabic{equation}}
\renewcommand{\thesection}{S\arabic{section}}
\renewcommand{\thefigure}{S\arabic{figure}}
\renewcommand{\thetable}{S\arabic{table}}
\setcounter{table}{0}
\setcounter{figure}{0}
\setcounter{equation}{0}
\setcounter{section}{0}

\begin{center}
\textbf{ \large{ Supplementary Information}} 
\end{center}
\tableofcontents

\clearpage
\section{ Experimental setup} 
	
\begin {figure}[!ht]
	\begin{center}
		\includegraphics[width=0.65\textwidth]{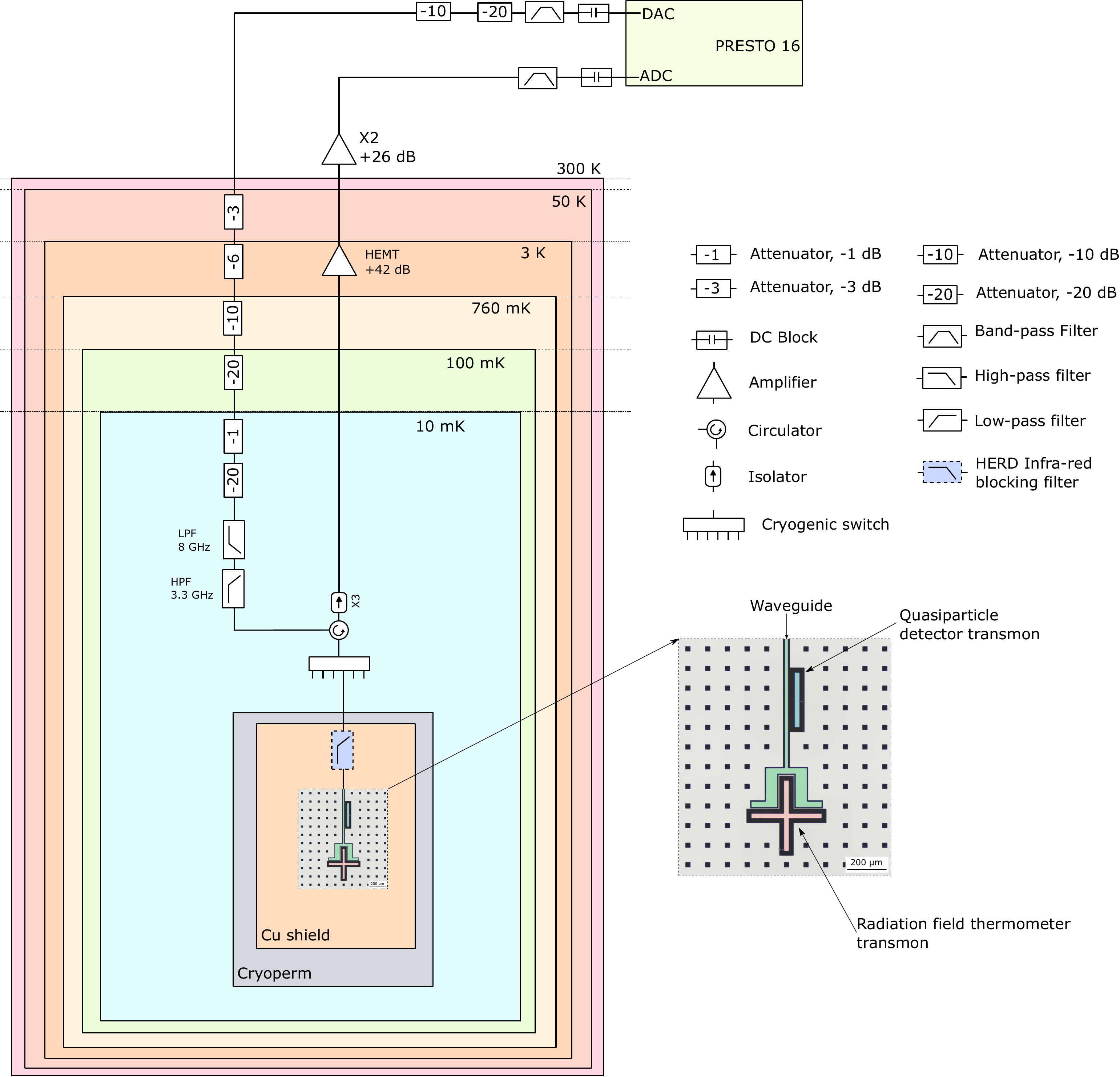}
		\caption{  Measurement setup.  Schematic diagram of the measurement setup, showing the measurement circuitry both inside and outside of the cryostat. A picture of the chip shown in the inset, marks the quasiparticle detector transmon coupled to the waveguide, and the rafiation-field thermometer~\cite{Scigliuzzo2020} coupled to the open end of the waveguide.
		\label{fig:setup}}	
	\end{center}
\end {figure}

	We measure the devices in a dilution refrigerator with a base temperature $T_{mix}\leq10$~mK. In order to achieve a low quasiparticle (QP) tunneling rate ($\Gamma_p$), suppressing the flux of high-energy radiation to to sample is important~\cite{Gordon2022}. We use radiation shields anchored to 50~K, 3~K, still (760~mK), and mixing chamber (10~mK) stages of the dilution refrigerator.  We further use a cryoperm shield and Cu shield, anchored to themixing chamber (10~mK) stage  to enclose the sample holder~\cite{202Qlab2023} containing the device. Furthermore, a magnetic shield is anchored to the outer vacuum can of the cryostat. We attenuate the input lines by using attenuators thermalized at different stages of the cryostat [Fig.~\ref{fig:setup}]. 
	
~\\ In the cooldown during which the data in Figs.~1-3 where taken, we mounted an additional high-energy radiation drain (HERD) filter~\cite{Rehammar2023} providing an attenuation in excess of 60~dB above 70~GHz, inside the cryoperm shield. However, the data presented in Fig.~4 were taken during a different cooldown, in which we did not use the filter. In both cases, we measured a QP tunneling rate $\Gamma_p$ between 2 and $4~\rm{s}^{-1}$, measured at the magic power. (Based on the data at hand, we cannot determine whether using the filter improves transmon-state-independent QP tunneling rate $\Gamma_0$ in our setup.)

~\\ From the fact that the addition of an IR-blocking filter did not improve $\Gamma_p$, we speculate that, in the present setup, $\Gamma_p$ is not limited by high-energy photons in the microwave lines. Indeed, in Ref.~\cite{Connolly2023}, it was found that the QP tunneling rate was sensitive to IR attenuation (provided by Eccosorb filters of various lengths),  only after an indium seal was added to the sample package. We note that the sensor presented here facilitates a detailed study of the impact of IR blocking filters on QP tunneling rates; such a study is, however, beyond the scope of this paper.

~\\ To generate the microwave tones and measurement, we use a microwave transceiver,  Presto~\cite{Tholen2022}, which performs direct digital synthesis of microwave signals using multiple Nyquist bands. We band-pass filter the generated signals to reject spurious signals at other Nyquist bands. We use the continuous-wave firmware (lock-in module) of Presto to perform the time-series measurement detecting QP tunneling events. To reduce the data rate, the signal is integrated over time bins of $1~\mu\rm{s}$.
We amplify the output signal by using a high electron mobility transistor (HEMT) amplifier at the 3~K stage, followed by two low-noise amplifiers at room temperature, before digitizing using the analog-to-digital converter of Presto.
	
\section{Device fabrication }
\label{sec:fab}
	
We fabricate the devices on 330~$\mu$m thick C-axis sapphire substrate. To fabricate the ground plane, we first deposit 150~nm Al by electron-beam evaporation. We use optical lithography and wet etching to pattern the ground plane, coplanar waveguide, and transmon pads. We subsequently fabricate Josephson junctions using electron-beam lithography, double-angle Al electron-beam evaporation (nominal thicknesses 50~nm and 110~nm, deposited at an angle of 45$^\circ$), in-situ oxidation, and liftoff. The junction design follows the Manhattan technique. 
	
\section{Micrograph of the full device}
	
\begin {figure}[!ht]
	\begin{center}
		\includegraphics[width=0.70\textwidth]{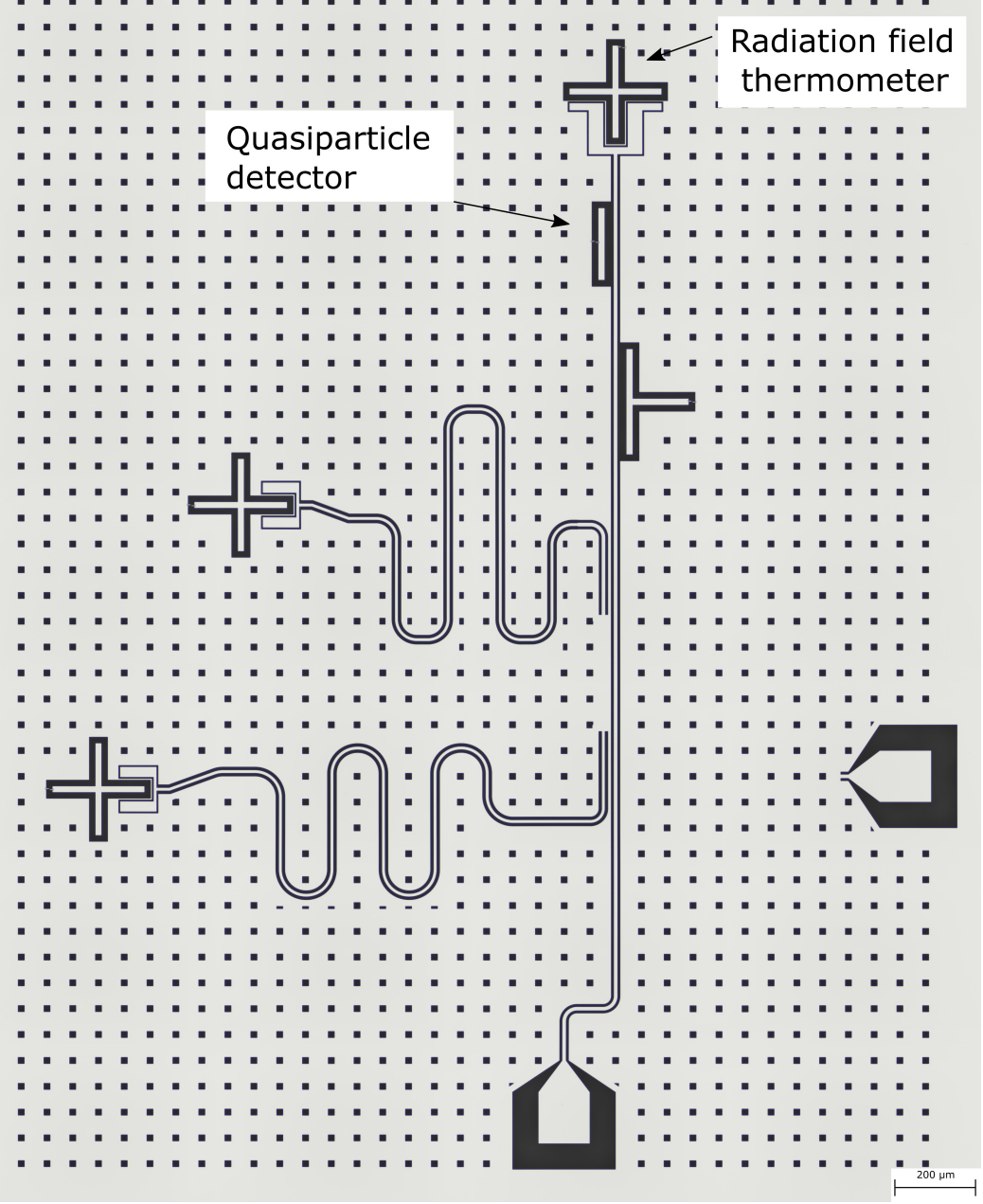}
		\caption{  Picture of the measured device.
			\label{fig:chip}}	
	\end{center}
\end {figure}

\section{ QP tunneling rate of multiple devices}
	
\begin {figure}[!ht]
	\begin{center}
		\includegraphics[width=0.95\textwidth]{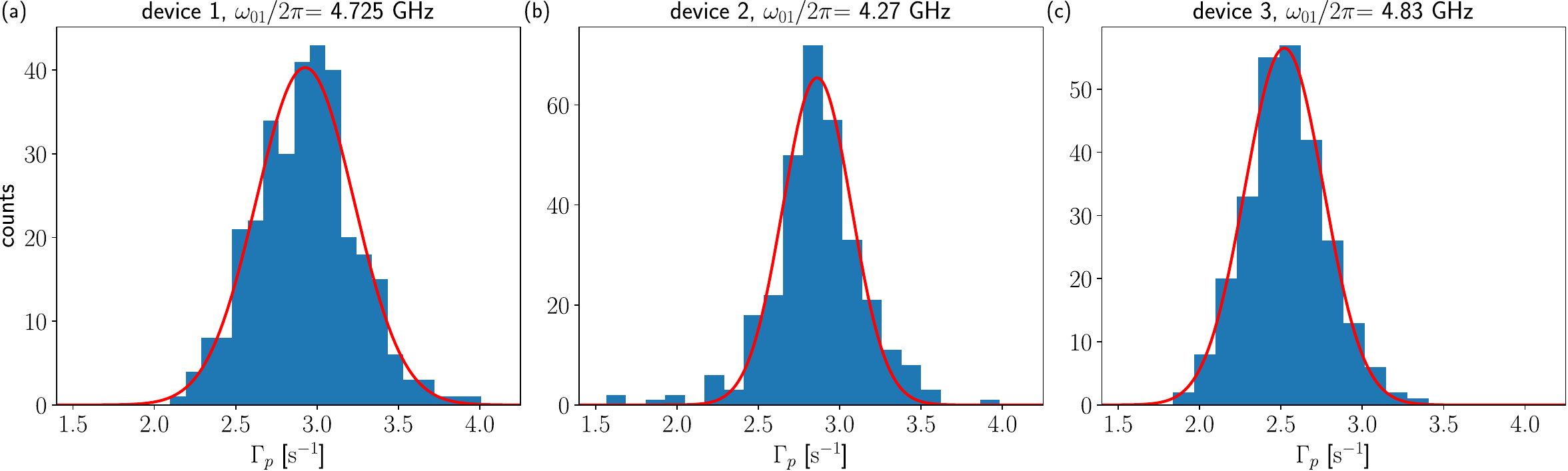}
		\caption{    Histogram of  $\Gamma_p$, for three devices, from data collected following a protocol described in Fig.~3(b) of the main text.
			\label{fig:sup_histo_qp_3dev}}	
	\end{center}
\end {figure}

In Fig.~\ref{fig:sup_histo_qp_3dev}, we show a histogram of $\Gamma_p$, measured for three devices. ``Device 1'' corresponds to the device for  which the data is presented in the manuscript.  The other two devices were fabricated in a different fabrication round and were measured in a separate cooldown, using the same measurement setup. The red lines are the Gaussian fit to the data. For all devices, the average rate is between 2.5 and 3~$\rm{s}^{-1}$.
	
\section{ Comparison of measured quasiparticle tunneling rates}

In Table~\ref{tbl:compare}, we list  quasiparticle tunneling rate $\Gamma_p$ reported over the past 6 years, measured using  Al-based charge-sensitive transmons, utilizing different device architectures and detection schemes.
	
\begin{center}
  \begin{table}[!ht]
    \caption{ Measured QP tunneling rates in Aluminum-based, charge-sensitive transmons  \label{tbl:compare} }
    \begin{tabular}{ p{0.5cm}  p{2.75cm} p{3.5cm} p{3.75cm}  p{3.5cm}  p{1.5cm} p{1.25cm}    }
      \hline
      \hline
      No. & Ref.  & Device & Architecture & Detection scheme & $\Gamma_p$~$[\rm{s}^{-1}]$ & Year   \\[0.1cm] \hline
      \hline
       1 & Serniak \textit{et. al.}~\cite{Serniak2018} &  transmon (split) & 3D &  Ramsey & 13000  & 2018   \\ \hline
       2 & Serniak \textit{et. al.}~\cite{Serniak2019} &  transmon (split) & 3D &  Direct dispersive readout & 166  & 2019   \\ \hline
	 3 &  Diamond \textit{et. al.}~\cite{Diamond2022} &  transmon (split) & 3D &  Ramsey & 340  & 2022   \\ \hline  
	 4 &  Gordon \textit{et. al.}~\cite{Gordon2022} &  transmon (split) & Planar & Ramsey & 6-10  &  2022  \\ \hline
	5  &   Liu \textit{et. al.}~\cite{Liu2022} &  transmon (single island) & Planar & Ramsey  &  13 &  2022  \\ \hline
	6  &   Pan \textit{et. al.}~\cite{Pan2022} & transmon (split) & Planar  & Ramsey & 0.5  & 2022   \\ \hline
	 7 &   Iaia \textit{et. al.}~\cite{Iaia2022} & transmon (single island) & Planar + phonon traps  & Ramsey & 0.023  & 2022   \\ \hline
	  8&   Kurter \textit{et. al.}~\cite{Kurter2022} & transmon (split) & Planar  & Ramsey & 70-1000  & 2022   \\ \hline
	9  &   Connolly \textit{et. al.}~\cite{Connolly2023} & transmon (split) &  3D  &  Direct dispersive readout  & 0.14  & 2023   \\ \hline 
	10&   Krause \textit{et. al.}~\cite{Krause2024} &  transmon (split)   & 3D + in-plane magnetic field & Ramsey  & 830-1250 & 2024  \\ \hline 
     11 & This work &    transmon (single island) & Planar & Direct scattering &  2-4  & 2024  \\ \hline
      \hline
    \end{tabular}
  \end{table}
\end{center}

\section{ Analysis of drive dependent QP tunneling rates }

\begin {figure}[!ht]
	\begin{center}
	\includegraphics[width=0.5\textwidth]{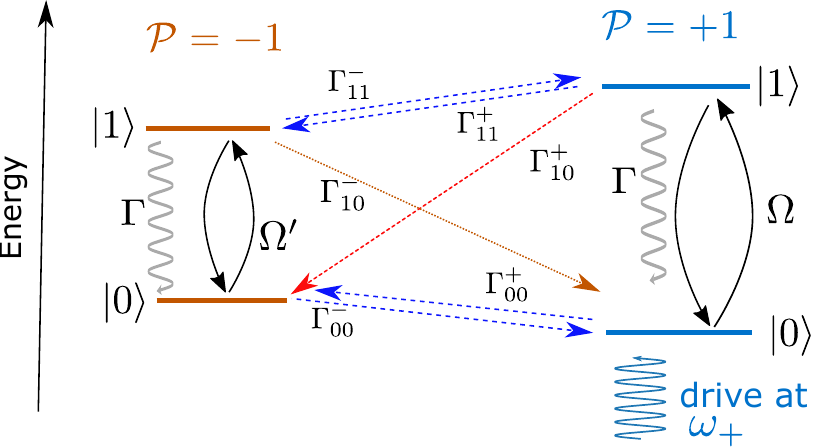}
	\caption{    Two charge parity states of a transmon approximated as a two-level system. The arrows mark allowed transitions. 
	\label{fig:sup_rates}}	
	\end{center}
\end {figure}

In order to understand the  drive power dependent QP tunneling rates and parity biasing [Fig.~4 in main text], we solve for drive-power dependent steady-state populations in the ground ($|0\rangle$) and the first excited state $|1\rangle$. We approximate the transmon as a two-level system with two parity states [Fig.~\ref{fig:sup_rates}].  We  write a rate equation of the form \begin{equation}
\frac{d\vec{p}}{dt}  = A \vec{p},
\label{eqn:rate}
\end{equation}
where the population vector $\vec{p}$ is written as
\begin{equation}
\vec{p} = \begin{pmatrix}  p_{0,+} \\    p_{1,+} \\ p_{0,-} \\ p_{1,-} \end{pmatrix}
\label{eqn:p}
\end{equation} 
Here,   $0$ and $1$ indicate the ground and first excited states, and $+$ and $-$ indicate the even and odd parity states, and $p$ denote population at each state; thus $p_{1,+}$ is the population of the excited state with parity $\mathcal{P}=+1$. 

~\\
Drive-induced population exchange between $|0\rangle$ and $|1\rangle$ states within same parity, spontaneous emission into the waveguide, and parity switching due to QP tunneling process are the competing processes that determine the steady state populations. The transitions we considered in the model are marked by arrows in Fig.~\ref{fig:sup_rates}. Together with the assumptions discussed below, they lead to  the rate matrix:
\begin{equation}
A=   \left(\begin{array}{cccc} - \big(\frac{\Omega^2}{\Gamma}  + \Gamma_0 \big) & 		\frac{\Omega^2}{\Gamma}  + \Gamma 			& \Gamma_0 		&  \Gamma_1 \\ 
							\frac{\Omega^2}{\Gamma}  & -( \frac{\Omega^2}{\Gamma}  +  \Gamma_0+ \Gamma_1+ \Gamma)	 & 0 & \Gamma_0 \\ 
							\Gamma_0 	& \Gamma_1 			& -\big( \frac{\Gamma \Omega^2}{\Gamma^2 + 4 \delta^2} + \Gamma_0 \big) 				& 	\frac{\Gamma \Omega^2}{\Gamma^2 + 4 \delta^2} + \Gamma \\ 
							0	 		& \Gamma_0 			& \frac{\Gamma \Omega^2}{\Gamma^2 + 4 \delta^2}  		& -\big( \frac{\Gamma \Omega^2}{\Gamma^2 + 4 \delta^2} + \Gamma_0 + \Gamma_1+\Gamma \big) \\ \end{array}\right) 
\label{eqn:mat}
\end{equation}

To arrive at this expression of $A$, we first note that the readout tone induces a population exchange rate:
\begin{equation}
\tilde{\Omega}(\delta) = \frac{\Gamma \Omega^2}{\Gamma^2 + 4 \delta^2},
\end{equation}
where $\Gamma$ is the decay rate of the transmon to the waveguide and  $\delta$ is the frequency detuning between the drive and the transmon parity state.  Thus $\delta=0$ for $\mathcal{P}=+1$ parity and $\delta=\omega_+ - \omega_-$ for the $\mathcal{P}=-1$ parity state. Here $\Omega$ is the drive strength, related to the input power $P_{\rm{in}}$  by the expression $\Omega = 2 \sqrt{\Gamma P_{\rm{in}} / \hbar \omega_{01} }$.
We have neglected nonradiative decay of the transmon $\Gamma_{\rm nr} \approx 2\pi \cdot 5~\rm{kHz} \ll \Gamma$ and thermal excitation in the waveguide, $n_{\rm th}\approx 0.001$ \cite{Scigliuzzo2020}, at the frequency of the transmon.

~\\
Regarding QP tunneling rates, we further assume that
\begin{itemize}
    \item QP tunneling rates are parity independent: $\Gamma^+_{00} = \Gamma^-_{00}$,  $\Gamma^+_{11} = \Gamma^-_{11}$, and $\Gamma^+_{10} = \Gamma^-_{10} = \Gamma_1$.
    \item Tunneling processes that preserve the state of the transmon do not depend on the transmon state: $\Gamma^\pm_{11} = \Gamma^\pm_{00}$.
    \item QP tuneling processes that drives a $|0\rangle \to |1\rangle$ transition are energetically suppressed. 
\end{itemize}
These assumptions are consistent with transmon-state-dependent  tunneling  of quasiparticles,  trapped in the low-superconducting gap  electrodes  of the transmon, where energy of the excited state of the transmon provide additional energy in their Arrhenius-type activated tunneling process~\cite{Connolly2023,Houzet2019}. Regions with different superconducting gaps within the transmon is a consequence of different thicknesses of Al films involved in the device fabrication process[see Sec.~\ref{sec:fab}]~\cite{Marchegiani2022}. 
We represent the qubit excitation preserving parity switching rates by $\Gamma_0 = \Gamma^+_{11} = \Gamma^-_{11} = \Gamma^+_{00} = \Gamma^-_{00}$, whereas any excess rate for $|1\rangle$ state is absorbed into the definition of $\Gamma_1$.

~\\
We solve for steady state of the coupled rate equations [Eqn.~\ref{eqn:rate}, with Eqn.~\ref{eqn:mat}] and obtain  average QP tunneling rate $\Gamma_p = \Gamma_0 + (p_{1,+} + p_{1,-})\Gamma_1$: 
\begin{equation}
\Gamma_p =  \frac{\left(\Gamma _0+\left(2 \Gamma _0+\Gamma _1\right) x^2\right) \left(4 \Gamma _0 \delta ^2+\Gamma ^2 \left(\Gamma _0+\left(2 \Gamma _0+\Gamma _1\right) x^2\right)\right)}{\Gamma _1 x^2 \left(2 \delta ^2+\Gamma ^2 \left(2 x^2+1\right)\right)+\Gamma _0 \left(2 x^2+1\right) \left(4 \delta ^2+\Gamma ^2 \left(2 x^2+1\right)\right)}
\label{eqn:qpt_rate}
\end{equation}
Here $x = \Omega/\Gamma$. We have used Eqn.~\ref{eqn:qpt_rate} to fit the drive power dependent $\Gamma_p$ in Fig.~3(a) of the main text, with  $\Gamma_0$ and $\Gamma_1$ being the only fit parameters, while $\delta$ and $\Gamma$ are obtained from independent characterization of the device.

~\\
The expectation value of parity $\langle \mathcal{P} \rangle$ is further calculated from the steady-state solution as $\langle \mathcal{P} \rangle = -1 (p_{0,-} + p_{1,-}) + (p_{0,+} + p_{1,+})$.


\begin{thebibliography}{61}%
\makeatletter
\providecommand \@ifxundefined [1]{%
 \@ifx{#1\undefined}
}%
\providecommand \@ifnum [1]{%
 \ifnum #1\expandafter \@firstoftwo
 \else \expandafter \@secondoftwo
 \fi
}%
\providecommand \@ifx [1]{%
 \ifx #1\expandafter \@firstoftwo
 \else \expandafter \@secondoftwo
 \fi
}%
\providecommand \natexlab [1]{#1}%
\providecommand \enquote  [1]{``#1''}%
\providecommand \bibnamefont  [1]{#1}%
\providecommand \bibfnamefont [1]{#1}%
\providecommand \citenamefont [1]{#1}%
\providecommand \href@noop [0]{\@secondoftwo}%
\providecommand \href [0]{\begingroup \@sanitize@url \@href}%
\providecommand \@href[1]{\@@startlink{#1}\@@href}%
\providecommand \@@href[1]{\endgroup#1\@@endlink}%
\providecommand \@sanitize@url [0]{\catcode `\\12\catcode `\$12\catcode
  `\&12\catcode `\#12\catcode `\^12\catcode `\_12\catcode `\%12\relax}%
\providecommand \@@startlink[1]{}%
\providecommand \@@endlink[0]{}%
\providecommand \url  [0]{\begingroup\@sanitize@url \@url }%
\providecommand \@url [1]{\endgroup\@href {#1}{\urlprefix }}%
\providecommand \urlprefix  [0]{URL }%
\providecommand \Eprint [0]{\href }%
\providecommand \doibase [0]{https://doi.org/}%
\providecommand \selectlanguage [0]{\@gobble}%
\providecommand \bibinfo  [0]{\@secondoftwo}%
\providecommand \bibfield  [0]{\@secondoftwo}%
\providecommand \translation [1]{[#1]}%
\providecommand \BibitemOpen [0]{}%
\providecommand \bibitemStop [0]{}%
\providecommand \bibitemNoStop [0]{.\EOS\space}%
\providecommand \EOS [0]{\spacefactor3000\relax}%
\providecommand \BibitemShut  [1]{\csname bibitem#1\endcsname}%
\let\auto@bib@innerbib\@empty
\bibitem [{\citenamefont {Glazman}\ and\ \citenamefont
  {Catelani}(2021)}]{Glazman2021}%
  \BibitemOpen
  \bibfield  {author} {\bibinfo {author} {\bibfnamefont {L.~I.}\ \bibnamefont
  {Glazman}}\ and\ \bibinfo {author} {\bibfnamefont {G.}~\bibnamefont
  {Catelani}},\ }\href {https://doi.org/10.21468/SciPostPhysLectNotes.31}
  {\bibfield  {journal} {\bibinfo  {journal} {SciPost Phys. Lect. Notes}\ ,\
  \bibinfo {pages} {31}} (\bibinfo {year} {2021})}\BibitemShut {NoStop}%
\bibitem [{\citenamefont {Segall}\ \emph {et~al.}(2004)\citenamefont {Segall},
  \citenamefont {Wilson}, \citenamefont {Li}, \citenamefont {Frunzio},
  \citenamefont {Friedrich}, \citenamefont {Gaidis},\ and\ \citenamefont
  {Prober}}]{Segall2004}%
  \BibitemOpen
  \bibfield  {author} {\bibinfo {author} {\bibfnamefont {K.}~\bibnamefont
  {Segall}}, \bibinfo {author} {\bibfnamefont {C.}~\bibnamefont {Wilson}},
  \bibinfo {author} {\bibfnamefont {L.}~\bibnamefont {Li}}, \bibinfo {author}
  {\bibfnamefont {L.}~\bibnamefont {Frunzio}}, \bibinfo {author} {\bibfnamefont
  {S.}~\bibnamefont {Friedrich}}, \bibinfo {author} {\bibfnamefont {M.~C.}\
  \bibnamefont {Gaidis}},\ and\ \bibinfo {author} {\bibfnamefont {D.~E.}\
  \bibnamefont {Prober}},\ }\href {https://doi.org/10.1103/PhysRevB.70.214520}
  {\bibfield  {journal} {\bibinfo  {journal} {Phys. Rev. B}\ }\textbf {\bibinfo
  {volume} {70}},\ \bibinfo {pages} {214520} (\bibinfo {year}
  {2004})}\BibitemShut {NoStop}%
\bibitem [{\citenamefont {Wang}\ \emph {et~al.}(2014)\citenamefont {Wang},
  \citenamefont {Gao}, \citenamefont {Pop}, \citenamefont {Vool}, \citenamefont
  {Axline}, \citenamefont {Brecht}, \citenamefont {Heeres}, \citenamefont
  {Frunzio}, \citenamefont {Devoret}, \citenamefont {Catelani}, \citenamefont
  {Glazman},\ and\ \citenamefont {Schoelkopf}}]{Wang2014}%
  \BibitemOpen
  \bibfield  {author} {\bibinfo {author} {\bibfnamefont {C.}~\bibnamefont
  {Wang}}, \bibinfo {author} {\bibfnamefont {Y.~Y.}\ \bibnamefont {Gao}},
  \bibinfo {author} {\bibfnamefont {I.~M.}\ \bibnamefont {Pop}}, \bibinfo
  {author} {\bibfnamefont {U.}~\bibnamefont {Vool}}, \bibinfo {author}
  {\bibfnamefont {C.}~\bibnamefont {Axline}}, \bibinfo {author} {\bibfnamefont
  {T.}~\bibnamefont {Brecht}}, \bibinfo {author} {\bibfnamefont {R.~W.}\
  \bibnamefont {Heeres}}, \bibinfo {author} {\bibfnamefont {L.}~\bibnamefont
  {Frunzio}}, \bibinfo {author} {\bibfnamefont {M.~H.}\ \bibnamefont
  {Devoret}}, \bibinfo {author} {\bibfnamefont {G.}~\bibnamefont {Catelani}},
  \bibinfo {author} {\bibfnamefont {L.~I.}\ \bibnamefont {Glazman}},\ and\
  \bibinfo {author} {\bibfnamefont {R.~J.}\ \bibnamefont {Schoelkopf}},\ }\href
  {https://doi.org/10.1038/ncomms6836} {\bibfield  {journal} {\bibinfo
  {journal} {Nature Communications}\ }\textbf {\bibinfo {volume} {5}},\
  \bibinfo {pages} {5836} (\bibinfo {year} {2014})}\BibitemShut {NoStop}%
\bibitem [{\citenamefont {de~Visser}\ \emph {et~al.}(2014)\citenamefont
  {de~Visser}, \citenamefont {Goldie}, \citenamefont {Diener}, \citenamefont
  {Withington}, \citenamefont {Baselmans},\ and\ \citenamefont
  {Klapwijk}}]{Visser2014}%
  \BibitemOpen
  \bibfield  {author} {\bibinfo {author} {\bibfnamefont {P.~J.}\ \bibnamefont
  {de~Visser}}, \bibinfo {author} {\bibfnamefont {D.~J.}\ \bibnamefont
  {Goldie}}, \bibinfo {author} {\bibfnamefont {P.}~\bibnamefont {Diener}},
  \bibinfo {author} {\bibfnamefont {S.}~\bibnamefont {Withington}}, \bibinfo
  {author} {\bibfnamefont {J.~J.~A.}\ \bibnamefont {Baselmans}},\ and\ \bibinfo
  {author} {\bibfnamefont {T.~M.}\ \bibnamefont {Klapwijk}},\ }\href
  {https://doi.org/10.1103/PhysRevLett.112.047004} {\bibfield  {journal}
  {\bibinfo  {journal} {Phys. Rev. Lett.}\ }\textbf {\bibinfo {volume} {112}},\
  \bibinfo {pages} {047004} (\bibinfo {year} {2014})}\BibitemShut {NoStop}%
\bibitem [{\citenamefont {Gr\"unhaupt}\ \emph {et~al.}(2018)\citenamefont
  {Gr\"unhaupt}, \citenamefont {Maleeva}, \citenamefont {Skacel}, \citenamefont
  {Calvo}, \citenamefont {Levy-Bertrand}, \citenamefont {Ustinov},
  \citenamefont {Rotzinger}, \citenamefont {Monfardini}, \citenamefont
  {Catelani},\ and\ \citenamefont {Pop}}]{Gruenhaupt2018}%
  \BibitemOpen
  \bibfield  {author} {\bibinfo {author} {\bibfnamefont {L.}~\bibnamefont
  {Gr\"unhaupt}}, \bibinfo {author} {\bibfnamefont {N.}~\bibnamefont
  {Maleeva}}, \bibinfo {author} {\bibfnamefont {S.~T.}\ \bibnamefont {Skacel}},
  \bibinfo {author} {\bibfnamefont {M.}~\bibnamefont {Calvo}}, \bibinfo
  {author} {\bibfnamefont {F.}~\bibnamefont {Levy-Bertrand}}, \bibinfo {author}
  {\bibfnamefont {A.~V.}\ \bibnamefont {Ustinov}}, \bibinfo {author}
  {\bibfnamefont {H.}~\bibnamefont {Rotzinger}}, \bibinfo {author}
  {\bibfnamefont {A.}~\bibnamefont {Monfardini}}, \bibinfo {author}
  {\bibfnamefont {G.}~\bibnamefont {Catelani}},\ and\ \bibinfo {author}
  {\bibfnamefont {I.~M.}\ \bibnamefont {Pop}},\ }\href
  {https://doi.org/10.1103/PhysRevLett.121.117001} {\bibfield  {journal}
  {\bibinfo  {journal} {Phys. Rev. Lett.}\ }\textbf {\bibinfo {volume} {121}},\
  \bibinfo {pages} {117001} (\bibinfo {year} {2018})}\BibitemShut {NoStop}%
\bibitem [{\citenamefont {Mannila}\ \emph {et~al.}(2022)\citenamefont
  {Mannila}, \citenamefont {Samuelsson}, \citenamefont {Simbierowicz},
  \citenamefont {Peltonen}, \citenamefont {Vesterinen}, \citenamefont
  {Grönberg}, \citenamefont {Hassel}, \citenamefont {Maisi},\ and\
  \citenamefont {Pekola}}]{Mannila2022}%
  \BibitemOpen
  \bibfield  {author} {\bibinfo {author} {\bibfnamefont {E.~T.}\ \bibnamefont
  {Mannila}}, \bibinfo {author} {\bibfnamefont {P.}~\bibnamefont {Samuelsson}},
  \bibinfo {author} {\bibfnamefont {S.}~\bibnamefont {Simbierowicz}}, \bibinfo
  {author} {\bibfnamefont {J.~T.}\ \bibnamefont {Peltonen}}, \bibinfo {author}
  {\bibfnamefont {V.}~\bibnamefont {Vesterinen}}, \bibinfo {author}
  {\bibfnamefont {L.}~\bibnamefont {Grönberg}}, \bibinfo {author}
  {\bibfnamefont {J.}~\bibnamefont {Hassel}}, \bibinfo {author} {\bibfnamefont
  {V.~F.}\ \bibnamefont {Maisi}},\ and\ \bibinfo {author} {\bibfnamefont
  {J.~P.}\ \bibnamefont {Pekola}},\ }\href
  {https://doi.org/10.1038/s41567-021-01433-7} {\bibfield  {journal} {\bibinfo
  {journal} {Nature Physics}\ }\textbf {\bibinfo {volume} {18}},\ \bibinfo
  {pages} {145} (\bibinfo {year} {2022})}\BibitemShut {NoStop}%
\bibitem [{\citenamefont {Pan}\ \emph {et~al.}(2022)\citenamefont {Pan},
  \citenamefont {Zhou}, \citenamefont {Yuan}, \citenamefont {Nie},
  \citenamefont {Wei}, \citenamefont {Zhang}, \citenamefont {Li}, \citenamefont
  {Liu}, \citenamefont {Jiang}, \citenamefont {Catelani}, \citenamefont {Hu},
  \citenamefont {Yan},\ and\ \citenamefont {Yu}}]{Pan2022}%
  \BibitemOpen
  \bibfield  {author} {\bibinfo {author} {\bibfnamefont {X.}~\bibnamefont
  {Pan}}, \bibinfo {author} {\bibfnamefont {Y.}~\bibnamefont {Zhou}}, \bibinfo
  {author} {\bibfnamefont {H.}~\bibnamefont {Yuan}}, \bibinfo {author}
  {\bibfnamefont {L.}~\bibnamefont {Nie}}, \bibinfo {author} {\bibfnamefont
  {W.}~\bibnamefont {Wei}}, \bibinfo {author} {\bibfnamefont {L.}~\bibnamefont
  {Zhang}}, \bibinfo {author} {\bibfnamefont {J.}~\bibnamefont {Li}}, \bibinfo
  {author} {\bibfnamefont {S.}~\bibnamefont {Liu}}, \bibinfo {author}
  {\bibfnamefont {Z.~H.}\ \bibnamefont {Jiang}}, \bibinfo {author}
  {\bibfnamefont {G.}~\bibnamefont {Catelani}}, \bibinfo {author}
  {\bibfnamefont {L.}~\bibnamefont {Hu}}, \bibinfo {author} {\bibfnamefont
  {F.}~\bibnamefont {Yan}},\ and\ \bibinfo {author} {\bibfnamefont
  {D.}~\bibnamefont {Yu}},\ }\href {https://doi.org/10.1038/s41467-022-34727-2}
  {\bibfield  {journal} {\bibinfo  {journal} {Nature Communications}\ }\textbf
  {\bibinfo {volume} {13}},\ \bibinfo {pages} {7196} (\bibinfo {year}
  {2022})}\BibitemShut {NoStop}%
\bibitem [{\citenamefont {Henriques}\ \emph {et~al.}(2019)\citenamefont
  {Henriques}, \citenamefont {Valenti}, \citenamefont {Charpentier},
  \citenamefont {Lagoin}, \citenamefont {Gouriou}, \citenamefont {Martínez},
  \citenamefont {Cardani}, \citenamefont {Vignati}, \citenamefont {Grünhaupt},
  \citenamefont {Gusenkova}, \citenamefont {Ferrero}, \citenamefont {Skacel},
  \citenamefont {Wernsdorfer}, \citenamefont {Ustinov}, \citenamefont
  {Catelani}, \citenamefont {Sander},\ and\ \citenamefont
  {Pop}}]{Henriques2019}%
  \BibitemOpen
  \bibfield  {author} {\bibinfo {author} {\bibfnamefont {F.}~\bibnamefont
  {Henriques}}, \bibinfo {author} {\bibfnamefont {F.}~\bibnamefont {Valenti}},
  \bibinfo {author} {\bibfnamefont {T.}~\bibnamefont {Charpentier}}, \bibinfo
  {author} {\bibfnamefont {M.}~\bibnamefont {Lagoin}}, \bibinfo {author}
  {\bibfnamefont {C.}~\bibnamefont {Gouriou}}, \bibinfo {author} {\bibfnamefont
  {M.}~\bibnamefont {Martínez}}, \bibinfo {author} {\bibfnamefont
  {L.}~\bibnamefont {Cardani}}, \bibinfo {author} {\bibfnamefont
  {M.}~\bibnamefont {Vignati}}, \bibinfo {author} {\bibfnamefont
  {L.}~\bibnamefont {Grünhaupt}}, \bibinfo {author} {\bibfnamefont
  {D.}~\bibnamefont {Gusenkova}}, \bibinfo {author} {\bibfnamefont
  {J.}~\bibnamefont {Ferrero}}, \bibinfo {author} {\bibfnamefont {S.~T.}\
  \bibnamefont {Skacel}}, \bibinfo {author} {\bibfnamefont {W.}~\bibnamefont
  {Wernsdorfer}}, \bibinfo {author} {\bibfnamefont {A.~V.}\ \bibnamefont
  {Ustinov}}, \bibinfo {author} {\bibfnamefont {G.}~\bibnamefont {Catelani}},
  \bibinfo {author} {\bibfnamefont {O.}~\bibnamefont {Sander}},\ and\ \bibinfo
  {author} {\bibfnamefont {I.~M.}\ \bibnamefont {Pop}},\ }\href
  {https://doi.org/10.1063/1.5124967} {\bibfield  {journal} {\bibinfo
  {journal} {Applied Physics Letters}\ }\textbf {\bibinfo {volume} {115}},\
  \bibinfo {pages} {212601} (\bibinfo {year} {2019})}\BibitemShut {NoStop}%
\bibitem [{\citenamefont {Cardani}\ \emph {et~al.}(2021)\citenamefont
  {Cardani}, \citenamefont {Valenti}, \citenamefont {Casali}, \citenamefont
  {Catelani}, \citenamefont {Charpentier}, \citenamefont {Clemenza},
  \citenamefont {Colantoni}, \citenamefont {Cruciani}, \citenamefont
  {D’Imperio}, \citenamefont {Gironi}, \citenamefont {Grünhaupt},
  \citenamefont {Gusenkova}, \citenamefont {Henriques}, \citenamefont {Lagoin},
  \citenamefont {Martinez}, \citenamefont {Pettinari}, \citenamefont {Rusconi},
  \citenamefont {Sander}, \citenamefont {Tomei}, \citenamefont {Ustinov},
  \citenamefont {Weber}, \citenamefont {Wernsdorfer}, \citenamefont {Vignati},
  \citenamefont {Pirro},\ and\ \citenamefont {Pop}}]{Cardani2021}%
  \BibitemOpen
  \bibfield  {author} {\bibinfo {author} {\bibfnamefont {L.}~\bibnamefont
  {Cardani}}, \bibinfo {author} {\bibfnamefont {F.}~\bibnamefont {Valenti}},
  \bibinfo {author} {\bibfnamefont {N.}~\bibnamefont {Casali}}, \bibinfo
  {author} {\bibfnamefont {G.}~\bibnamefont {Catelani}}, \bibinfo {author}
  {\bibfnamefont {T.}~\bibnamefont {Charpentier}}, \bibinfo {author}
  {\bibfnamefont {M.}~\bibnamefont {Clemenza}}, \bibinfo {author}
  {\bibfnamefont {I.}~\bibnamefont {Colantoni}}, \bibinfo {author}
  {\bibfnamefont {A.}~\bibnamefont {Cruciani}}, \bibinfo {author}
  {\bibfnamefont {G.}~\bibnamefont {D’Imperio}}, \bibinfo {author}
  {\bibfnamefont {L.}~\bibnamefont {Gironi}}, \bibinfo {author} {\bibfnamefont
  {L.}~\bibnamefont {Grünhaupt}}, \bibinfo {author} {\bibfnamefont
  {D.}~\bibnamefont {Gusenkova}}, \bibinfo {author} {\bibfnamefont
  {F.}~\bibnamefont {Henriques}}, \bibinfo {author} {\bibfnamefont
  {M.}~\bibnamefont {Lagoin}}, \bibinfo {author} {\bibfnamefont
  {M.}~\bibnamefont {Martinez}}, \bibinfo {author} {\bibfnamefont
  {G.}~\bibnamefont {Pettinari}}, \bibinfo {author} {\bibfnamefont
  {C.}~\bibnamefont {Rusconi}}, \bibinfo {author} {\bibfnamefont
  {O.}~\bibnamefont {Sander}}, \bibinfo {author} {\bibfnamefont
  {C.}~\bibnamefont {Tomei}}, \bibinfo {author} {\bibfnamefont {A.~V.}\
  \bibnamefont {Ustinov}}, \bibinfo {author} {\bibfnamefont {M.}~\bibnamefont
  {Weber}}, \bibinfo {author} {\bibfnamefont {W.}~\bibnamefont {Wernsdorfer}},
  \bibinfo {author} {\bibfnamefont {M.}~\bibnamefont {Vignati}}, \bibinfo
  {author} {\bibfnamefont {S.}~\bibnamefont {Pirro}},\ and\ \bibinfo {author}
  {\bibfnamefont {I.~M.}\ \bibnamefont {Pop}},\ }\href
  {https://doi.org/10.1038/s41467-021-23032-z} {\bibfield  {journal} {\bibinfo
  {journal} {Nature Communications}\ }\textbf {\bibinfo {volume} {12}},\
  \bibinfo {pages} {2733} (\bibinfo {year} {2021})}\BibitemShut {NoStop}%
\bibitem [{\citenamefont {Iaia}\ \emph {et~al.}(2022)\citenamefont {Iaia},
  \citenamefont {Ku}, \citenamefont {Ballard}, \citenamefont {Larson},
  \citenamefont {Yelton}, \citenamefont {Liu}, \citenamefont {Patel},
  \citenamefont {McDermott},\ and\ \citenamefont {Plourde}}]{Iaia2022}%
  \BibitemOpen
  \bibfield  {author} {\bibinfo {author} {\bibfnamefont {V.}~\bibnamefont
  {Iaia}}, \bibinfo {author} {\bibfnamefont {J.}~\bibnamefont {Ku}}, \bibinfo
  {author} {\bibfnamefont {A.}~\bibnamefont {Ballard}}, \bibinfo {author}
  {\bibfnamefont {C.~P.}\ \bibnamefont {Larson}}, \bibinfo {author}
  {\bibfnamefont {E.}~\bibnamefont {Yelton}}, \bibinfo {author} {\bibfnamefont
  {C.~H.}\ \bibnamefont {Liu}}, \bibinfo {author} {\bibfnamefont
  {S.}~\bibnamefont {Patel}}, \bibinfo {author} {\bibfnamefont
  {R.}~\bibnamefont {McDermott}},\ and\ \bibinfo {author} {\bibfnamefont
  {B.~L.~T.}\ \bibnamefont {Plourde}},\ }\href
  {https://doi.org/10.1038/s41467-022-33997-0} {\bibfield  {journal} {\bibinfo
  {journal} {Nature Communications}\ }\textbf {\bibinfo {volume} {13}},\
  \bibinfo {pages} {6425} (\bibinfo {year} {2022})}\BibitemShut {NoStop}%
\bibitem [{\citenamefont {Wilen}\ \emph {et~al.}(2021)\citenamefont {Wilen},
  \citenamefont {Abdullah}, \citenamefont {Kurinsky}, \citenamefont {Stanford},
  \citenamefont {Cardani}, \citenamefont {D’Imperio}, \citenamefont {Tomei},
  \citenamefont {Faoro}, \citenamefont {Ioffe}, \citenamefont {Liu},
  \citenamefont {Opremcak}, \citenamefont {Christensen}, \citenamefont
  {DuBois},\ and\ \citenamefont {McDermott}}]{Wilen2021}%
  \BibitemOpen
  \bibfield  {author} {\bibinfo {author} {\bibfnamefont {C.~D.}\ \bibnamefont
  {Wilen}}, \bibinfo {author} {\bibfnamefont {S.}~\bibnamefont {Abdullah}},
  \bibinfo {author} {\bibfnamefont {N.~A.}\ \bibnamefont {Kurinsky}}, \bibinfo
  {author} {\bibfnamefont {C.}~\bibnamefont {Stanford}}, \bibinfo {author}
  {\bibfnamefont {L.}~\bibnamefont {Cardani}}, \bibinfo {author} {\bibfnamefont
  {G.}~\bibnamefont {D’Imperio}}, \bibinfo {author} {\bibfnamefont
  {C.}~\bibnamefont {Tomei}}, \bibinfo {author} {\bibfnamefont
  {L.}~\bibnamefont {Faoro}}, \bibinfo {author} {\bibfnamefont {L.~B.}\
  \bibnamefont {Ioffe}}, \bibinfo {author} {\bibfnamefont {C.~H.}\ \bibnamefont
  {Liu}}, \bibinfo {author} {\bibfnamefont {A.}~\bibnamefont {Opremcak}},
  \bibinfo {author} {\bibfnamefont {B.~G.}\ \bibnamefont {Christensen}},
  \bibinfo {author} {\bibfnamefont {J.~L.}\ \bibnamefont {DuBois}},\ and\
  \bibinfo {author} {\bibfnamefont {R.}~\bibnamefont {McDermott}},\ }\href
  {https://doi.org/10.1038/s41586-021-03557-5} {\bibfield  {journal} {\bibinfo
  {journal} {Nature}\ }\textbf {\bibinfo {volume} {594}},\ \bibinfo {pages}
  {369} (\bibinfo {year} {2021})}\BibitemShut {NoStop}%
\bibitem [{\citenamefont {Martinis}(2021)}]{Martinis2021}%
  \BibitemOpen
  \bibfield  {author} {\bibinfo {author} {\bibfnamefont {J.~M.}\ \bibnamefont
  {Martinis}},\ }\href {https://doi.org/10.1038/s41534-021-00431-0} {\bibfield
  {journal} {\bibinfo  {journal} {npj Quantum Information}\ }\textbf {\bibinfo
  {volume} {7}},\ \bibinfo {pages} {90} (\bibinfo {year} {2021})}\BibitemShut
  {NoStop}%
\bibitem [{\citenamefont {Jalabert}\ \emph {et~al.}(2023)\citenamefont
  {Jalabert}, \citenamefont {Driessen}, \citenamefont {Gustavo}, \citenamefont
  {Thomassin}, \citenamefont {Levy-Bertrand},\ and\ \citenamefont
  {Chapelier}}]{Jalabert2023}%
  \BibitemOpen
  \bibfield  {author} {\bibinfo {author} {\bibfnamefont {T.}~\bibnamefont
  {Jalabert}}, \bibinfo {author} {\bibfnamefont {E.~F.~C.}\ \bibnamefont
  {Driessen}}, \bibinfo {author} {\bibfnamefont {F.}~\bibnamefont {Gustavo}},
  \bibinfo {author} {\bibfnamefont {J.~L.}\ \bibnamefont {Thomassin}}, \bibinfo
  {author} {\bibfnamefont {F.}~\bibnamefont {Levy-Bertrand}},\ and\ \bibinfo
  {author} {\bibfnamefont {C.}~\bibnamefont {Chapelier}},\ }\href
  {https://doi.org/10.1038/s41567-023-01999-4} {\bibfield  {journal} {\bibinfo
  {journal} {Nature Physics}\ }\textbf {\bibinfo {volume} {19}},\ \bibinfo
  {pages} {956} (\bibinfo {year} {2023})}\BibitemShut {NoStop}%
\bibitem [{\citenamefont {Gr{\"u}nhaupt}\ \emph {et~al.}(2019)\citenamefont
  {Gr{\"u}nhaupt}, \citenamefont {Spiecker}, \citenamefont {Gusenkova},
  \citenamefont {Maleeva}, \citenamefont {Skacel}, \citenamefont {Takmakov},
  \citenamefont {Valenti}, \citenamefont {Winkel}, \citenamefont {Rotzinger},
  \citenamefont {Wernsdorfer}, \citenamefont {Ustinov},\ and\ \citenamefont
  {Pop}}]{Gruenhaupt2019}%
  \BibitemOpen
  \bibfield  {author} {\bibinfo {author} {\bibfnamefont {L.}~\bibnamefont
  {Gr{\"u}nhaupt}}, \bibinfo {author} {\bibfnamefont {M.}~\bibnamefont
  {Spiecker}}, \bibinfo {author} {\bibfnamefont {D.}~\bibnamefont {Gusenkova}},
  \bibinfo {author} {\bibfnamefont {N.}~\bibnamefont {Maleeva}}, \bibinfo
  {author} {\bibfnamefont {S.~T.}\ \bibnamefont {Skacel}}, \bibinfo {author}
  {\bibfnamefont {I.}~\bibnamefont {Takmakov}}, \bibinfo {author}
  {\bibfnamefont {F.}~\bibnamefont {Valenti}}, \bibinfo {author} {\bibfnamefont
  {P.}~\bibnamefont {Winkel}}, \bibinfo {author} {\bibfnamefont
  {H.}~\bibnamefont {Rotzinger}}, \bibinfo {author} {\bibfnamefont
  {W.}~\bibnamefont {Wernsdorfer}}, \bibinfo {author} {\bibfnamefont {A.~V.}\
  \bibnamefont {Ustinov}},\ and\ \bibinfo {author} {\bibfnamefont {I.~M.}\
  \bibnamefont {Pop}},\ }\href
  {https://www.nature.com/articles/s41563-019-0350-3} {\bibfield  {journal}
  {\bibinfo  {journal} {Nature Materials}\ } (\bibinfo {year}
  {2019})}\BibitemShut {NoStop}%
\bibitem [{\citenamefont {Amin}\ \emph {et~al.}(2022)\citenamefont {Amin},
  \citenamefont {Ladner}, \citenamefont {Jourdan}, \citenamefont {Hentz},
  \citenamefont {Roch},\ and\ \citenamefont {Renard}}]{Amin2022}%
  \BibitemOpen
  \bibfield  {author} {\bibinfo {author} {\bibfnamefont {K.~R.}\ \bibnamefont
  {Amin}}, \bibinfo {author} {\bibfnamefont {C.}~\bibnamefont {Ladner}},
  \bibinfo {author} {\bibfnamefont {G.}~\bibnamefont {Jourdan}}, \bibinfo
  {author} {\bibfnamefont {S.}~\bibnamefont {Hentz}}, \bibinfo {author}
  {\bibfnamefont {N.}~\bibnamefont {Roch}},\ and\ \bibinfo {author}
  {\bibfnamefont {J.}~\bibnamefont {Renard}},\ }\href
  {https://doi.org/10.1063/5.0086019} {\bibfield  {journal} {\bibinfo
  {journal} {Applied Physics Letters}\ }\textbf {\bibinfo {volume} {120}},\
  \bibinfo {pages} {164001} (\bibinfo {year} {2022})}\BibitemShut {NoStop}%
\bibitem [{\citenamefont {Siddiqi}(2021)}]{Siddiqi2021}%
  \BibitemOpen
  \bibfield  {author} {\bibinfo {author} {\bibfnamefont {I.}~\bibnamefont
  {Siddiqi}},\ }\href {https://doi.org/10.1038/s41578-021-00370-4} {\bibfield
  {journal} {\bibinfo  {journal} {Nature Reviews Materials}\ }\textbf {\bibinfo
  {volume} {6}},\ \bibinfo {pages} {875} (\bibinfo {year} {2021})}\BibitemShut
  {NoStop}%
\bibitem [{\citenamefont {Aumentado}\ \emph {et~al.}(2023)\citenamefont
  {Aumentado}, \citenamefont {Catelani},\ and\ \citenamefont
  {Serniak}}]{Aumentado2023}%
  \BibitemOpen
  \bibfield  {author} {\bibinfo {author} {\bibfnamefont {J.}~\bibnamefont
  {Aumentado}}, \bibinfo {author} {\bibfnamefont {G.}~\bibnamefont
  {Catelani}},\ and\ \bibinfo {author} {\bibfnamefont {K.}~\bibnamefont
  {Serniak}},\ }\href {https://doi.org/10.1063/PT.3.5291} {\bibfield  {journal}
  {\bibinfo  {journal} {Physics Today}\ }\textbf {\bibinfo {volume} {76}},\
  \bibinfo {pages} {34} (\bibinfo {year} {2023})}\BibitemShut {NoStop}%
\bibitem [{\citenamefont {Gusenkova}\ \emph {et~al.}(2022)\citenamefont
  {Gusenkova}, \citenamefont {Valenti}, \citenamefont {Spiecker}, \citenamefont
  {Günzler}, \citenamefont {Paluch}, \citenamefont {Rieger}, \citenamefont
  {Pioraş-Ţimbolmaş}, \citenamefont {Zârbo}, \citenamefont {Casali},
  \citenamefont {Colantoni}, \citenamefont {Cruciani}, \citenamefont {Pirro},
  \citenamefont {Cardani}, \citenamefont {Petrescu}, \citenamefont
  {Wernsdorfer}, \citenamefont {Winkel},\ and\ \citenamefont
  {Pop}}]{Gusenkova2022}%
  \BibitemOpen
  \bibfield  {author} {\bibinfo {author} {\bibfnamefont {D.}~\bibnamefont
  {Gusenkova}}, \bibinfo {author} {\bibfnamefont {F.}~\bibnamefont {Valenti}},
  \bibinfo {author} {\bibfnamefont {M.}~\bibnamefont {Spiecker}}, \bibinfo
  {author} {\bibfnamefont {S.}~\bibnamefont {Günzler}}, \bibinfo {author}
  {\bibfnamefont {P.}~\bibnamefont {Paluch}}, \bibinfo {author} {\bibfnamefont
  {D.}~\bibnamefont {Rieger}}, \bibinfo {author} {\bibfnamefont {L.-M.}\
  \bibnamefont {Pioraş-Ţimbolmaş}}, \bibinfo {author} {\bibfnamefont
  {L.~P.}\ \bibnamefont {Zârbo}}, \bibinfo {author} {\bibfnamefont
  {N.}~\bibnamefont {Casali}}, \bibinfo {author} {\bibfnamefont
  {I.}~\bibnamefont {Colantoni}}, \bibinfo {author} {\bibfnamefont
  {A.}~\bibnamefont {Cruciani}}, \bibinfo {author} {\bibfnamefont
  {S.}~\bibnamefont {Pirro}}, \bibinfo {author} {\bibfnamefont
  {L.}~\bibnamefont {Cardani}}, \bibinfo {author} {\bibfnamefont
  {A.}~\bibnamefont {Petrescu}}, \bibinfo {author} {\bibfnamefont
  {W.}~\bibnamefont {Wernsdorfer}}, \bibinfo {author} {\bibfnamefont
  {P.}~\bibnamefont {Winkel}},\ and\ \bibinfo {author} {\bibfnamefont {I.~M.}\
  \bibnamefont {Pop}},\ }\href {https://doi.org/10.1063/5.0075909} {\bibfield
  {journal} {\bibinfo  {journal} {Applied Physics Letters}\ }\textbf {\bibinfo
  {volume} {120}},\ \bibinfo {pages} {054001} (\bibinfo {year} {2022})},\
  \Eprint
  {https://arxiv.org/abs/https://pubs.aip.org/aip/apl/article-pdf/doi/10.1063/5.0075909/16476464/054001\_1\_online.pdf}
  {https://pubs.aip.org/aip/apl/article-pdf/doi/10.1063/5.0075909/16476464/054001\_1\_online.pdf}
  \BibitemShut {NoStop}%
\bibitem [{\citenamefont {Pop}\ \emph {et~al.}(2014)\citenamefont {Pop},
  \citenamefont {Geerlings}, \citenamefont {Catelani}, \citenamefont
  {Schoelkopf}, \citenamefont {Glazman},\ and\ \citenamefont
  {Devoret}}]{Pop2014}%
  \BibitemOpen
  \bibfield  {author} {\bibinfo {author} {\bibfnamefont {I.~M.}\ \bibnamefont
  {Pop}}, \bibinfo {author} {\bibfnamefont {K.}~\bibnamefont {Geerlings}},
  \bibinfo {author} {\bibfnamefont {G.}~\bibnamefont {Catelani}}, \bibinfo
  {author} {\bibfnamefont {R.~J.}\ \bibnamefont {Schoelkopf}}, \bibinfo
  {author} {\bibfnamefont {L.~I.}\ \bibnamefont {Glazman}},\ and\ \bibinfo
  {author} {\bibfnamefont {M.~H.}\ \bibnamefont {Devoret}},\ }\href
  {https://doi.org/10.1038/nature13017} {\bibfield  {journal} {\bibinfo
  {journal} {Nature}\ }\textbf {\bibinfo {volume} {508}},\ \bibinfo {pages}
  {369} (\bibinfo {year} {2014})}\BibitemShut {NoStop}%
\bibitem [{\citenamefont {Shaw}\ \emph {et~al.}(2009)\citenamefont {Shaw},
  \citenamefont {Bueno}, \citenamefont {Day}, \citenamefont {Bradford},\ and\
  \citenamefont {Echternach}}]{Shaw2009}%
  \BibitemOpen
  \bibfield  {author} {\bibinfo {author} {\bibfnamefont {M.~D.}\ \bibnamefont
  {Shaw}}, \bibinfo {author} {\bibfnamefont {J.}~\bibnamefont {Bueno}},
  \bibinfo {author} {\bibfnamefont {P.}~\bibnamefont {Day}}, \bibinfo {author}
  {\bibfnamefont {C.~M.}\ \bibnamefont {Bradford}},\ and\ \bibinfo {author}
  {\bibfnamefont {P.~M.}\ \bibnamefont {Echternach}},\ }\href
  {https://doi.org/10.1103/PhysRevB.79.144511} {\bibfield  {journal} {\bibinfo
  {journal} {Phys. Rev. B}\ }\textbf {\bibinfo {volume} {79}},\ \bibinfo
  {pages} {144511} (\bibinfo {year} {2009})}\BibitemShut {NoStop}%
\bibitem [{\citenamefont {Bockstiegel}\ \emph {et~al.}(2014)\citenamefont
  {Bockstiegel}, \citenamefont {Gao}, \citenamefont {Vissers}, \citenamefont
  {Sandberg}, \citenamefont {Chaudhuri}, \citenamefont {Sanders}, \citenamefont
  {Vale}, \citenamefont {Irwin},\ and\ \citenamefont
  {Pappas}}]{bockstiegel2014}%
  \BibitemOpen
  \bibfield  {author} {\bibinfo {author} {\bibfnamefont {C.}~\bibnamefont
  {Bockstiegel}}, \bibinfo {author} {\bibfnamefont {J.}~\bibnamefont {Gao}},
  \bibinfo {author} {\bibfnamefont {M.~R.}\ \bibnamefont {Vissers}}, \bibinfo
  {author} {\bibfnamefont {M.}~\bibnamefont {Sandberg}}, \bibinfo {author}
  {\bibfnamefont {S.}~\bibnamefont {Chaudhuri}}, \bibinfo {author}
  {\bibfnamefont {A.}~\bibnamefont {Sanders}}, \bibinfo {author} {\bibfnamefont
  {L.~R.}\ \bibnamefont {Vale}}, \bibinfo {author} {\bibfnamefont {K.~D.}\
  \bibnamefont {Irwin}},\ and\ \bibinfo {author} {\bibfnamefont {D.~P.}\
  \bibnamefont {Pappas}},\ }\href {https://doi.org/10.1007/s10909-013-1042-z}
  {\bibfield  {journal} {\bibinfo  {journal} {Journal of Low Temperature
  Physics}\ }\textbf {\bibinfo {volume} {176}},\ \bibinfo {pages} {476}
  (\bibinfo {year} {2014})}\BibitemShut {NoStop}%
\bibitem [{\citenamefont {Esmaeil~Zadeh}\ \emph {et~al.}(2021)\citenamefont
  {Esmaeil~Zadeh}, \citenamefont {Chang}, \citenamefont {Los}, \citenamefont
  {Gyger}, \citenamefont {Elshaari}, \citenamefont {Steinhauer}, \citenamefont
  {Dorenbos},\ and\ \citenamefont {Zwiller}}]{EsmaeilZadeh2021}%
  \BibitemOpen
  \bibfield  {author} {\bibinfo {author} {\bibfnamefont {I.}~\bibnamefont
  {Esmaeil~Zadeh}}, \bibinfo {author} {\bibfnamefont {J.}~\bibnamefont
  {Chang}}, \bibinfo {author} {\bibfnamefont {J.~W.~N.}\ \bibnamefont {Los}},
  \bibinfo {author} {\bibfnamefont {S.}~\bibnamefont {Gyger}}, \bibinfo
  {author} {\bibfnamefont {A.~W.}\ \bibnamefont {Elshaari}}, \bibinfo {author}
  {\bibfnamefont {S.}~\bibnamefont {Steinhauer}}, \bibinfo {author}
  {\bibfnamefont {S.~N.}\ \bibnamefont {Dorenbos}},\ and\ \bibinfo {author}
  {\bibfnamefont {V.}~\bibnamefont {Zwiller}},\ }\href
  {https://doi.org/10.1063/5.0045990} {\bibfield  {journal} {\bibinfo
  {journal} {Applied Physics Letters}\ }\textbf {\bibinfo {volume} {118}},\
  \bibinfo {pages} {190502} (\bibinfo {year} {2021})}\BibitemShut {NoStop}%
\bibitem [{\citenamefont {Naaman}\ and\ \citenamefont
  {Aumentado}(2006)}]{Naaman2006}%
  \BibitemOpen
  \bibfield  {author} {\bibinfo {author} {\bibfnamefont {O.}~\bibnamefont
  {Naaman}}\ and\ \bibinfo {author} {\bibfnamefont {J.}~\bibnamefont
  {Aumentado}},\ }\href {https://doi.org/10.1103/PhysRevB.73.172504} {\bibfield
   {journal} {\bibinfo  {journal} {Phys. Rev. B}\ }\textbf {\bibinfo {volume}
  {73}},\ \bibinfo {pages} {172504} (\bibinfo {year} {2006})}\BibitemShut
  {NoStop}%
\bibitem [{\citenamefont {Ferguson}\ \emph {et~al.}(2006)\citenamefont
  {Ferguson}, \citenamefont {Court}, \citenamefont {Hudson},\ and\
  \citenamefont {Clark}}]{Ferguson2006}%
  \BibitemOpen
  \bibfield  {author} {\bibinfo {author} {\bibfnamefont {A.~J.}\ \bibnamefont
  {Ferguson}}, \bibinfo {author} {\bibfnamefont {N.~A.}\ \bibnamefont {Court}},
  \bibinfo {author} {\bibfnamefont {F.~E.}\ \bibnamefont {Hudson}},\ and\
  \bibinfo {author} {\bibfnamefont {R.~G.}\ \bibnamefont {Clark}},\ }\href@noop
  {} {\bibfield  {journal} {\bibinfo  {journal} {Phys. Rev. Lett.}\ }\textbf
  {\bibinfo {volume} {97}},\ \bibinfo {pages} {106603} (\bibinfo {year}
  {2006})}\BibitemShut {NoStop}%
\bibitem [{\citenamefont {Koch}\ \emph {et~al.}(2007)\citenamefont {Koch},
  \citenamefont {Yu}, \citenamefont {Gambetta}, \citenamefont {Houck},
  \citenamefont {Schuster}, \citenamefont {Majer}, \citenamefont {Blais},
  \citenamefont {Devoret}, \citenamefont {Girvin},\ and\ \citenamefont
  {Schoelkopf}}]{Koch2007}%
  \BibitemOpen
  \bibfield  {author} {\bibinfo {author} {\bibfnamefont {J.}~\bibnamefont
  {Koch}}, \bibinfo {author} {\bibfnamefont {T.~M.}\ \bibnamefont {Yu}},
  \bibinfo {author} {\bibfnamefont {J.}~\bibnamefont {Gambetta}}, \bibinfo
  {author} {\bibfnamefont {A.~A.}\ \bibnamefont {Houck}}, \bibinfo {author}
  {\bibfnamefont {D.~I.}\ \bibnamefont {Schuster}}, \bibinfo {author}
  {\bibfnamefont {J.}~\bibnamefont {Majer}}, \bibinfo {author} {\bibfnamefont
  {A.}~\bibnamefont {Blais}}, \bibinfo {author} {\bibfnamefont {M.~H.}\
  \bibnamefont {Devoret}}, \bibinfo {author} {\bibfnamefont {S.~M.}\
  \bibnamefont {Girvin}},\ and\ \bibinfo {author} {\bibfnamefont {R.~J.}\
  \bibnamefont {Schoelkopf}},\ }\href
  {https://doi.org/10.1103/PhysRevA.76.042319} {\bibfield  {journal} {\bibinfo
  {journal} {Phys. Rev. A}\ }\textbf {\bibinfo {volume} {76}},\ \bibinfo
  {pages} {042319} (\bibinfo {year} {2007})}\BibitemShut {NoStop}%
\bibitem [{\citenamefont {Ristè}\ \emph {et~al.}(2013)\citenamefont {Ristè},
  \citenamefont {Bultink}, \citenamefont {Tiggelman}, \citenamefont {Schouten},
  \citenamefont {Lehnert},\ and\ \citenamefont {DiCarlo}}]{Riste2013}%
  \BibitemOpen
  \bibfield  {author} {\bibinfo {author} {\bibfnamefont {D.}~\bibnamefont
  {Ristè}}, \bibinfo {author} {\bibfnamefont {C.~C.}\ \bibnamefont {Bultink}},
  \bibinfo {author} {\bibfnamefont {M.~J.}\ \bibnamefont {Tiggelman}}, \bibinfo
  {author} {\bibfnamefont {R.~N.}\ \bibnamefont {Schouten}}, \bibinfo {author}
  {\bibfnamefont {K.~W.}\ \bibnamefont {Lehnert}},\ and\ \bibinfo {author}
  {\bibfnamefont {L.}~\bibnamefont {DiCarlo}},\ }\href
  {https://doi.org/10.1038/ncomms2936} {\bibfield  {journal} {\bibinfo
  {journal} {Nature Communications}\ }\textbf {\bibinfo {volume} {4}},\
  \bibinfo {pages} {1913} (\bibinfo {year} {2013})}\BibitemShut {NoStop}%
\bibitem [{\citenamefont {Serniak}\ \emph {et~al.}(2018)\citenamefont
  {Serniak}, \citenamefont {Hays}, \citenamefont {de~Lange}, \citenamefont
  {Diamond}, \citenamefont {Shankar}, \citenamefont {Burkhart}, \citenamefont
  {Frunzio}, \citenamefont {Houzet},\ and\ \citenamefont
  {Devoret}}]{Serniak2018}%
  \BibitemOpen
  \bibfield  {author} {\bibinfo {author} {\bibfnamefont {K.}~\bibnamefont
  {Serniak}}, \bibinfo {author} {\bibfnamefont {M.}~\bibnamefont {Hays}},
  \bibinfo {author} {\bibfnamefont {G.}~\bibnamefont {de~Lange}}, \bibinfo
  {author} {\bibfnamefont {S.}~\bibnamefont {Diamond}}, \bibinfo {author}
  {\bibfnamefont {S.}~\bibnamefont {Shankar}}, \bibinfo {author} {\bibfnamefont
  {L.~D.}\ \bibnamefont {Burkhart}}, \bibinfo {author} {\bibfnamefont
  {L.}~\bibnamefont {Frunzio}}, \bibinfo {author} {\bibfnamefont
  {M.}~\bibnamefont {Houzet}},\ and\ \bibinfo {author} {\bibfnamefont {M.~H.}\
  \bibnamefont {Devoret}},\ }\href
  {https://doi.org/10.1103/PhysRevLett.121.157701} {\bibfield  {journal}
  {\bibinfo  {journal} {Phys. Rev. Lett.}\ }\textbf {\bibinfo {volume} {121}},\
  \bibinfo {pages} {157701} (\bibinfo {year} {2018})}\BibitemShut {NoStop}%
\bibitem [{\citenamefont {Serniak}\ \emph {et~al.}(2019)\citenamefont
  {Serniak}, \citenamefont {Diamond}, \citenamefont {Hays}, \citenamefont
  {Fatemi}, \citenamefont {Shankar}, \citenamefont {Frunzio}, \citenamefont
  {Schoelkopf},\ and\ \citenamefont {Devoret}}]{Serniak2019}%
  \BibitemOpen
  \bibfield  {author} {\bibinfo {author} {\bibfnamefont {K.}~\bibnamefont
  {Serniak}}, \bibinfo {author} {\bibfnamefont {S.}~\bibnamefont {Diamond}},
  \bibinfo {author} {\bibfnamefont {M.}~\bibnamefont {Hays}}, \bibinfo {author}
  {\bibfnamefont {V.}~\bibnamefont {Fatemi}}, \bibinfo {author} {\bibfnamefont
  {S.}~\bibnamefont {Shankar}}, \bibinfo {author} {\bibfnamefont
  {L.}~\bibnamefont {Frunzio}}, \bibinfo {author} {\bibfnamefont
  {R.}~\bibnamefont {Schoelkopf}},\ and\ \bibinfo {author} {\bibfnamefont
  {M.}~\bibnamefont {Devoret}},\ }\href
  {https://doi.org/10.1103/PhysRevApplied.12.014052} {\bibfield  {journal}
  {\bibinfo  {journal} {Phys. Rev. Appl.}\ }\textbf {\bibinfo {volume} {12}},\
  \bibinfo {pages} {014052} (\bibinfo {year} {2019})}\BibitemShut {NoStop}%
\bibitem [{\citenamefont {Connolly}\ \emph {et~al.}(2023)\citenamefont
  {Connolly}, \citenamefont {Kurilovich}, \citenamefont {Diamond},
  \citenamefont {Nho}, \citenamefont {Bøttcher}, \citenamefont {Glazman},
  \citenamefont {Fatemi},\ and\ \citenamefont {Devoret}}]{Connolly2023}%
  \BibitemOpen
  \bibfield  {author} {\bibinfo {author} {\bibfnamefont {T.}~\bibnamefont
  {Connolly}}, \bibinfo {author} {\bibfnamefont {P.~D.}\ \bibnamefont
  {Kurilovich}}, \bibinfo {author} {\bibfnamefont {S.}~\bibnamefont {Diamond}},
  \bibinfo {author} {\bibfnamefont {H.}~\bibnamefont {Nho}}, \bibinfo {author}
  {\bibfnamefont {C.~G.~L.}\ \bibnamefont {Bøttcher}}, \bibinfo {author}
  {\bibfnamefont {L.~I.}\ \bibnamefont {Glazman}}, \bibinfo {author}
  {\bibfnamefont {V.}~\bibnamefont {Fatemi}},\ and\ \bibinfo {author}
  {\bibfnamefont {M.~H.}\ \bibnamefont {Devoret}},\ }\href@noop {} {\bibinfo
  {title} {Coexistence of nonequilibrium density and equilibrium energy
  distribution of quasiparticles in a superconducting qubit}} (\bibinfo {year}
  {2023}),\ \Eprint {https://arxiv.org/abs/2302.12330} {arXiv:2302.12330
  [quant-ph]} \BibitemShut {NoStop}%
\bibitem [{\citenamefont {Diamond}\ \emph {et~al.}(2022)\citenamefont
  {Diamond}, \citenamefont {Fatemi}, \citenamefont {Hays}, \citenamefont {Nho},
  \citenamefont {Kurilovich}, \citenamefont {Connolly}, \citenamefont {Joshi},
  \citenamefont {Serniak}, \citenamefont {Frunzio}, \citenamefont {Glazman},\
  and\ \citenamefont {Devoret}}]{Diamond2022}%
  \BibitemOpen
  \bibfield  {author} {\bibinfo {author} {\bibfnamefont {S.}~\bibnamefont
  {Diamond}}, \bibinfo {author} {\bibfnamefont {V.}~\bibnamefont {Fatemi}},
  \bibinfo {author} {\bibfnamefont {M.}~\bibnamefont {Hays}}, \bibinfo {author}
  {\bibfnamefont {H.}~\bibnamefont {Nho}}, \bibinfo {author} {\bibfnamefont
  {P.~D.}\ \bibnamefont {Kurilovich}}, \bibinfo {author} {\bibfnamefont
  {T.}~\bibnamefont {Connolly}}, \bibinfo {author} {\bibfnamefont {V.~R.}\
  \bibnamefont {Joshi}}, \bibinfo {author} {\bibfnamefont {K.}~\bibnamefont
  {Serniak}}, \bibinfo {author} {\bibfnamefont {L.}~\bibnamefont {Frunzio}},
  \bibinfo {author} {\bibfnamefont {L.~I.}\ \bibnamefont {Glazman}},\ and\
  \bibinfo {author} {\bibfnamefont {M.~H.}\ \bibnamefont {Devoret}},\ }\href
  {https://doi.org/10.1103/PRXQuantum.3.040304} {\bibfield  {journal} {\bibinfo
   {journal} {PRX Quantum}\ }\textbf {\bibinfo {volume} {3}},\ \bibinfo {pages}
  {040304} (\bibinfo {year} {2022})}\BibitemShut {NoStop}%
\bibitem [{\citenamefont {Gordon}\ \emph {et~al.}(2022)\citenamefont {Gordon},
  \citenamefont {Murray}, \citenamefont {Kurter}, \citenamefont {Sandberg},
  \citenamefont {Hall}, \citenamefont {Balakrishnan}, \citenamefont {Shelby},
  \citenamefont {Wacaser}, \citenamefont {Stabile}, \citenamefont {Sleight},
  \citenamefont {Brink}, \citenamefont {Rothwell}, \citenamefont {Rodbell},
  \citenamefont {Dial},\ and\ \citenamefont {Steffen}}]{Gordon2022}%
  \BibitemOpen
  \bibfield  {author} {\bibinfo {author} {\bibfnamefont {R.~T.}\ \bibnamefont
  {Gordon}}, \bibinfo {author} {\bibfnamefont {C.~E.}\ \bibnamefont {Murray}},
  \bibinfo {author} {\bibfnamefont {C.}~\bibnamefont {Kurter}}, \bibinfo
  {author} {\bibfnamefont {M.}~\bibnamefont {Sandberg}}, \bibinfo {author}
  {\bibfnamefont {S.~A.}\ \bibnamefont {Hall}}, \bibinfo {author}
  {\bibfnamefont {K.}~\bibnamefont {Balakrishnan}}, \bibinfo {author}
  {\bibfnamefont {R.}~\bibnamefont {Shelby}}, \bibinfo {author} {\bibfnamefont
  {B.}~\bibnamefont {Wacaser}}, \bibinfo {author} {\bibfnamefont {A.~A.}\
  \bibnamefont {Stabile}}, \bibinfo {author} {\bibfnamefont {J.~W.}\
  \bibnamefont {Sleight}}, \bibinfo {author} {\bibfnamefont {M.}~\bibnamefont
  {Brink}}, \bibinfo {author} {\bibfnamefont {M.~B.}\ \bibnamefont {Rothwell}},
  \bibinfo {author} {\bibfnamefont {K.~P.}\ \bibnamefont {Rodbell}}, \bibinfo
  {author} {\bibfnamefont {O.}~\bibnamefont {Dial}},\ and\ \bibinfo {author}
  {\bibfnamefont {M.}~\bibnamefont {Steffen}},\ }\href
  {https://doi.org/10.1063/5.0078785} {\bibfield  {journal} {\bibinfo
  {journal} {Applied Physics Letters}\ }\textbf {\bibinfo {volume} {120}},\
  \bibinfo {pages} {074002} (\bibinfo {year} {2022})}\BibitemShut {NoStop}%
\bibitem [{\citenamefont {Liu}\ \emph {et~al.}(2022)\citenamefont {Liu},
  \citenamefont {Harrison}, \citenamefont {Patel}, \citenamefont {Wilen},
  \citenamefont {Rafferty}, \citenamefont {Shearrow}, \citenamefont {Ballard},
  \citenamefont {Iaia}, \citenamefont {Ku}, \citenamefont {Plourde},\ and\
  \citenamefont {McDermott}}]{Liu2022}%
  \BibitemOpen
  \bibfield  {author} {\bibinfo {author} {\bibfnamefont {C.-H.}\ \bibnamefont
  {Liu}}, \bibinfo {author} {\bibfnamefont {D.~C.}\ \bibnamefont {Harrison}},
  \bibinfo {author} {\bibfnamefont {S.}~\bibnamefont {Patel}}, \bibinfo
  {author} {\bibfnamefont {C.~D.}\ \bibnamefont {Wilen}}, \bibinfo {author}
  {\bibfnamefont {O.}~\bibnamefont {Rafferty}}, \bibinfo {author}
  {\bibfnamefont {A.}~\bibnamefont {Shearrow}}, \bibinfo {author}
  {\bibfnamefont {A.}~\bibnamefont {Ballard}}, \bibinfo {author} {\bibfnamefont
  {V.}~\bibnamefont {Iaia}}, \bibinfo {author} {\bibfnamefont {J.}~\bibnamefont
  {Ku}}, \bibinfo {author} {\bibfnamefont {B.~L.~T.}\ \bibnamefont {Plourde}},\
  and\ \bibinfo {author} {\bibfnamefont {R.}~\bibnamefont {McDermott}},\
  }\href@noop {} {\bibinfo {title} {Quasiparticle poisoning of superconducting
  qubits from resonant absorption of pair-breaking photons}} (\bibinfo {year}
  {2022}),\ \Eprint {https://arxiv.org/abs/2203.06577} {arXiv:2203.06577
  [quant-ph]} \BibitemShut {NoStop}%
\bibitem [{\citenamefont {Kamenov}\ \emph {et~al.}(2023)\citenamefont
  {Kamenov}, \citenamefont {DiNapoli}, \citenamefont {Gershenson},\ and\
  \citenamefont {Chakram}}]{Kamenov2023}%
  \BibitemOpen
  \bibfield  {author} {\bibinfo {author} {\bibfnamefont {P.}~\bibnamefont
  {Kamenov}}, \bibinfo {author} {\bibfnamefont {T.}~\bibnamefont {DiNapoli}},
  \bibinfo {author} {\bibfnamefont {M.}~\bibnamefont {Gershenson}},\ and\
  \bibinfo {author} {\bibfnamefont {S.}~\bibnamefont {Chakram}},\ }\href@noop
  {} {\bibinfo {title} {Suppression of quasiparticle poisoning in transmon
  qubits by gap engineering}} (\bibinfo {year} {2023}),\ \Eprint
  {https://arxiv.org/abs/2309.02655} {arXiv:2309.02655 [quant-ph]} \BibitemShut
  {NoStop}%
\bibitem [{\citenamefont {Chen}\ \emph {et~al.}(2023)\citenamefont {Chen},
  \citenamefont {Li}, \citenamefont {Lu}, \citenamefont {Warren}, \citenamefont
  {Križan}, \citenamefont {Kosen}, \citenamefont {Rommel}, \citenamefont
  {Ahmed}, \citenamefont {Osman}, \citenamefont {Biznárová}, \citenamefont
  {Fadavi~Roudsari}, \citenamefont {Lienhard}, \citenamefont {Caputo},
  \citenamefont {Grigoras}, \citenamefont {Grönberg}, \citenamefont
  {Govenius}, \citenamefont {Kockum}, \citenamefont {Delsing}, \citenamefont
  {Bylander},\ and\ \citenamefont {Tancredi}}]{Chen2023}%
  \BibitemOpen
  \bibfield  {author} {\bibinfo {author} {\bibfnamefont {L.}~\bibnamefont
  {Chen}}, \bibinfo {author} {\bibfnamefont {H.-X.}\ \bibnamefont {Li}},
  \bibinfo {author} {\bibfnamefont {Y.}~\bibnamefont {Lu}}, \bibinfo {author}
  {\bibfnamefont {C.~W.}\ \bibnamefont {Warren}}, \bibinfo {author}
  {\bibfnamefont {C.~J.}\ \bibnamefont {Križan}}, \bibinfo {author}
  {\bibfnamefont {S.}~\bibnamefont {Kosen}}, \bibinfo {author} {\bibfnamefont
  {M.}~\bibnamefont {Rommel}}, \bibinfo {author} {\bibfnamefont
  {S.}~\bibnamefont {Ahmed}}, \bibinfo {author} {\bibfnamefont
  {A.}~\bibnamefont {Osman}}, \bibinfo {author} {\bibfnamefont
  {J.}~\bibnamefont {Biznárová}}, \bibinfo {author} {\bibfnamefont
  {A.}~\bibnamefont {Fadavi~Roudsari}}, \bibinfo {author} {\bibfnamefont
  {B.}~\bibnamefont {Lienhard}}, \bibinfo {author} {\bibfnamefont
  {M.}~\bibnamefont {Caputo}}, \bibinfo {author} {\bibfnamefont
  {K.}~\bibnamefont {Grigoras}}, \bibinfo {author} {\bibfnamefont
  {L.}~\bibnamefont {Grönberg}}, \bibinfo {author} {\bibfnamefont
  {J.}~\bibnamefont {Govenius}}, \bibinfo {author} {\bibfnamefont {A.~F.}\
  \bibnamefont {Kockum}}, \bibinfo {author} {\bibfnamefont {P.}~\bibnamefont
  {Delsing}}, \bibinfo {author} {\bibfnamefont {J.}~\bibnamefont {Bylander}},\
  and\ \bibinfo {author} {\bibfnamefont {G.}~\bibnamefont {Tancredi}},\ }\href
  {https://doi.org/10.1038/s41534-023-00689-6} {\bibfield  {journal} {\bibinfo
  {journal} {npj Quantum Information}\ }\textbf {\bibinfo {volume} {9}},\
  \bibinfo {pages} {26} (\bibinfo {year} {2023})}\BibitemShut {NoStop}%
\bibitem [{\citenamefont {Astafiev}\ \emph {et~al.}(2010)\citenamefont
  {Astafiev}, \citenamefont {Zagoskin}, \citenamefont {Abdumalikov},
  \citenamefont {Pashkin}, \citenamefont {Yamamoto}, \citenamefont {Inomata},
  \citenamefont {Nakamura},\ and\ \citenamefont {Tsai}}]{Astafiev2010}%
  \BibitemOpen
  \bibfield  {author} {\bibinfo {author} {\bibfnamefont {O.}~\bibnamefont
  {Astafiev}}, \bibinfo {author} {\bibfnamefont {A.~M.}\ \bibnamefont
  {Zagoskin}}, \bibinfo {author} {\bibfnamefont {A.~A.}\ \bibnamefont
  {Abdumalikov}}, \bibinfo {author} {\bibfnamefont {Y.~A.}\ \bibnamefont
  {Pashkin}}, \bibinfo {author} {\bibfnamefont {T.}~\bibnamefont {Yamamoto}},
  \bibinfo {author} {\bibfnamefont {K.}~\bibnamefont {Inomata}}, \bibinfo
  {author} {\bibfnamefont {Y.}~\bibnamefont {Nakamura}},\ and\ \bibinfo
  {author} {\bibfnamefont {J.~S.}\ \bibnamefont {Tsai}},\ }\href
  {https://doi.org/10.1126/science.1181918} {\bibfield  {journal} {\bibinfo
  {journal} {Science}\ }\textbf {\bibinfo {volume} {327}},\ \bibinfo {pages}
  {840} (\bibinfo {year} {2010})},\ \Eprint
  {https://arxiv.org/abs/https://www.science.org/doi/pdf/10.1126/science.1181918}
  {https://www.science.org/doi/pdf/10.1126/science.1181918} \BibitemShut
  {NoStop}%
\bibitem [{\citenamefont {Winkel}\ \emph {et~al.}(2020)\citenamefont {Winkel},
  \citenamefont {Borisov}, \citenamefont {Gr\"unhaupt}, \citenamefont {Rieger},
  \citenamefont {Spiecker}, \citenamefont {Valenti}, \citenamefont {Ustinov},
  \citenamefont {Wernsdorfer},\ and\ \citenamefont {Pop}}]{Winkel2020a}%
  \BibitemOpen
  \bibfield  {author} {\bibinfo {author} {\bibfnamefont {P.}~\bibnamefont
  {Winkel}}, \bibinfo {author} {\bibfnamefont {K.}~\bibnamefont {Borisov}},
  \bibinfo {author} {\bibfnamefont {L.}~\bibnamefont {Gr\"unhaupt}}, \bibinfo
  {author} {\bibfnamefont {D.}~\bibnamefont {Rieger}}, \bibinfo {author}
  {\bibfnamefont {M.}~\bibnamefont {Spiecker}}, \bibinfo {author}
  {\bibfnamefont {F.}~\bibnamefont {Valenti}}, \bibinfo {author} {\bibfnamefont
  {A.~V.}\ \bibnamefont {Ustinov}}, \bibinfo {author} {\bibfnamefont
  {W.}~\bibnamefont {Wernsdorfer}},\ and\ \bibinfo {author} {\bibfnamefont
  {I.~M.}\ \bibnamefont {Pop}},\ }\href
  {https://doi.org/10.1103/PhysRevX.10.031032} {\bibfield  {journal} {\bibinfo
  {journal} {Phys. Rev. X}\ }\textbf {\bibinfo {volume} {10}},\ \bibinfo
  {pages} {031032} (\bibinfo {year} {2020})}\BibitemShut {NoStop}%
\bibitem [{\citenamefont {Aamir}\ \emph {et~al.}(2022)\citenamefont {Aamir},
  \citenamefont {Moreno}, \citenamefont {Sundelin}, \citenamefont
  {Bizn\'arov\'a}, \citenamefont {Scigliuzzo}, \citenamefont {Patel},
  \citenamefont {Osman}, \citenamefont {Lozano}, \citenamefont {Strandberg},\
  and\ \citenamefont {Gasparinetti}}]{Aamir2022}%
  \BibitemOpen
  \bibfield  {author} {\bibinfo {author} {\bibfnamefont {M.~A.}\ \bibnamefont
  {Aamir}}, \bibinfo {author} {\bibfnamefont {C.~C.}\ \bibnamefont {Moreno}},
  \bibinfo {author} {\bibfnamefont {S.}~\bibnamefont {Sundelin}}, \bibinfo
  {author} {\bibfnamefont {J.}~\bibnamefont {Bizn\'arov\'a}}, \bibinfo {author}
  {\bibfnamefont {M.}~\bibnamefont {Scigliuzzo}}, \bibinfo {author}
  {\bibfnamefont {K.~E.}\ \bibnamefont {Patel}}, \bibinfo {author}
  {\bibfnamefont {A.}~\bibnamefont {Osman}}, \bibinfo {author} {\bibfnamefont
  {D.~P.}\ \bibnamefont {Lozano}}, \bibinfo {author} {\bibfnamefont
  {I.}~\bibnamefont {Strandberg}},\ and\ \bibinfo {author} {\bibfnamefont
  {S.}~\bibnamefont {Gasparinetti}},\ }\href
  {https://doi.org/10.1103/PhysRevLett.129.123604} {\bibfield  {journal}
  {\bibinfo  {journal} {Phys. Rev. Lett.}\ }\textbf {\bibinfo {volume} {129}},\
  \bibinfo {pages} {123604} (\bibinfo {year} {2022})}\BibitemShut {NoStop}%
\bibitem [{\citenamefont {Fink}\ \emph {et~al.}(2023)\citenamefont {Fink},
  \citenamefont {Salemi}, \citenamefont {Young}, \citenamefont {Schuster},\
  and\ \citenamefont {Kurinsky}}]{Fink2023}%
  \BibitemOpen
  \bibfield  {author} {\bibinfo {author} {\bibfnamefont {C.~W.}\ \bibnamefont
  {Fink}}, \bibinfo {author} {\bibfnamefont {C.~P.}\ \bibnamefont {Salemi}},
  \bibinfo {author} {\bibfnamefont {B.~A.}\ \bibnamefont {Young}}, \bibinfo
  {author} {\bibfnamefont {D.~I.}\ \bibnamefont {Schuster}},\ and\ \bibinfo
  {author} {\bibfnamefont {N.~A.}\ \bibnamefont {Kurinsky}},\ }\href@noop {}
  {\bibinfo {title} {The superconducting quasiparticle-amplifying transmon: A
  qubit-based sensor for mev scale phonons and single thz photons}} (\bibinfo
  {year} {2023}),\ \Eprint {https://arxiv.org/abs/2310.01345} {arXiv:2310.01345
  [physics.ins-det]} \BibitemShut {NoStop}%
\bibitem [{\citenamefont {Scigliuzzo}\ \emph {et~al.}(2020)\citenamefont
  {Scigliuzzo}, \citenamefont {Bengtsson}, \citenamefont {Besse}, \citenamefont
  {Wallraff}, \citenamefont {Delsing},\ and\ \citenamefont
  {Gasparinetti}}]{Scigliuzzo2020}%
  \BibitemOpen
  \bibfield  {author} {\bibinfo {author} {\bibfnamefont {M.}~\bibnamefont
  {Scigliuzzo}}, \bibinfo {author} {\bibfnamefont {A.}~\bibnamefont
  {Bengtsson}}, \bibinfo {author} {\bibfnamefont {J.-C.}\ \bibnamefont
  {Besse}}, \bibinfo {author} {\bibfnamefont {A.}~\bibnamefont {Wallraff}},
  \bibinfo {author} {\bibfnamefont {P.}~\bibnamefont {Delsing}},\ and\ \bibinfo
  {author} {\bibfnamefont {S.}~\bibnamefont {Gasparinetti}},\ }\href
  {https://doi.org/10.1103/PhysRevX.10.041054} {\bibfield  {journal} {\bibinfo
  {journal} {Phys. Rev. X}\ }\textbf {\bibinfo {volume} {10}},\ \bibinfo
  {pages} {041054} (\bibinfo {year} {2020})}\BibitemShut {NoStop}%
\bibitem [{\citenamefont {Yuzhelevski}\ \emph {et~al.}(2000)\citenamefont
  {Yuzhelevski}, \citenamefont {Yuzhelevski},\ and\ \citenamefont
  {Jung}}]{Yuzhelevski2000}%
  \BibitemOpen
  \bibfield  {author} {\bibinfo {author} {\bibfnamefont {Y.}~\bibnamefont
  {Yuzhelevski}}, \bibinfo {author} {\bibfnamefont {M.}~\bibnamefont
  {Yuzhelevski}},\ and\ \bibinfo {author} {\bibfnamefont {G.}~\bibnamefont
  {Jung}},\ }\href {https://doi.org/10.1063/1.1150519} {\bibfield  {journal}
  {\bibinfo  {journal} {Review of Scientific Instruments}\ }\textbf {\bibinfo
  {volume} {71}},\ \bibinfo {pages} {1681} (\bibinfo {year}
  {2000})}\BibitemShut {NoStop}%
\bibitem [{\citenamefont {Schl\"or}\ \emph {et~al.}(2019)\citenamefont
  {Schl\"or}, \citenamefont {Lisenfeld}, \citenamefont {M\"uller},
  \citenamefont {Bilmes}, \citenamefont {Schneider}, \citenamefont {Pappas},
  \citenamefont {Ustinov},\ and\ \citenamefont {Weides}}]{Schloer2019}%
  \BibitemOpen
  \bibfield  {author} {\bibinfo {author} {\bibfnamefont {S.}~\bibnamefont
  {Schl\"or}}, \bibinfo {author} {\bibfnamefont {J.}~\bibnamefont {Lisenfeld}},
  \bibinfo {author} {\bibfnamefont {C.}~\bibnamefont {M\"uller}}, \bibinfo
  {author} {\bibfnamefont {A.}~\bibnamefont {Bilmes}}, \bibinfo {author}
  {\bibfnamefont {A.}~\bibnamefont {Schneider}}, \bibinfo {author}
  {\bibfnamefont {D.~P.}\ \bibnamefont {Pappas}}, \bibinfo {author}
  {\bibfnamefont {A.~V.}\ \bibnamefont {Ustinov}},\ and\ \bibinfo {author}
  {\bibfnamefont {M.}~\bibnamefont {Weides}},\ }\href
  {https://doi.org/10.1103/PhysRevLett.123.190502} {\bibfield  {journal}
  {\bibinfo  {journal} {Phys. Rev. Lett.}\ }\textbf {\bibinfo {volume} {123}},\
  \bibinfo {pages} {190502} (\bibinfo {year} {2019})}\BibitemShut {NoStop}%
\bibitem [{\citenamefont {Tennant}\ \emph {et~al.}(2022)\citenamefont
  {Tennant}, \citenamefont {Martinez}, \citenamefont {Beck}, \citenamefont
  {O'Kelley}, \citenamefont {Wilen}, \citenamefont {McDermott}, \citenamefont
  {DuBois},\ and\ \citenamefont {Rosen}}]{Tennant2022}%
  \BibitemOpen
  \bibfield  {author} {\bibinfo {author} {\bibfnamefont {D.~M.}\ \bibnamefont
  {Tennant}}, \bibinfo {author} {\bibfnamefont {L.~A.}\ \bibnamefont
  {Martinez}}, \bibinfo {author} {\bibfnamefont {K.~M.}\ \bibnamefont {Beck}},
  \bibinfo {author} {\bibfnamefont {S.~R.}\ \bibnamefont {O'Kelley}}, \bibinfo
  {author} {\bibfnamefont {C.~D.}\ \bibnamefont {Wilen}}, \bibinfo {author}
  {\bibfnamefont {R.}~\bibnamefont {McDermott}}, \bibinfo {author}
  {\bibfnamefont {J.~L.}\ \bibnamefont {DuBois}},\ and\ \bibinfo {author}
  {\bibfnamefont {Y.~J.}\ \bibnamefont {Rosen}},\ }\href
  {https://doi.org/10.1103/PRXQuantum.3.030307} {\bibfield  {journal} {\bibinfo
   {journal} {PRX Quantum}\ }\textbf {\bibinfo {volume} {3}},\ \bibinfo {pages}
  {030307} (\bibinfo {year} {2022})}\BibitemShut {NoStop}%
\bibitem [{\citenamefont {Damme}\ \emph {et~al.}(2024)\citenamefont {Damme},
  \citenamefont {Massar}, \citenamefont {Acharya}, \citenamefont {Ivanov},
  \citenamefont {Lozano}, \citenamefont {Canvel}, \citenamefont {Demarets},
  \citenamefont {Vangoidsenhoven}, \citenamefont {Hermans}, \citenamefont
  {Lai}, \citenamefont {Rao}, \citenamefont {Mongillo}, \citenamefont {Wan},
  \citenamefont {Boeck}, \citenamefont {Potocnik},\ and\ \citenamefont
  {Greve}}]{Damme2024}%
  \BibitemOpen
  \bibfield  {author} {\bibinfo {author} {\bibfnamefont {J.~V.}\ \bibnamefont
  {Damme}}, \bibinfo {author} {\bibfnamefont {S.}~\bibnamefont {Massar}},
  \bibinfo {author} {\bibfnamefont {R.}~\bibnamefont {Acharya}}, \bibinfo
  {author} {\bibfnamefont {T.}~\bibnamefont {Ivanov}}, \bibinfo {author}
  {\bibfnamefont {D.~P.}\ \bibnamefont {Lozano}}, \bibinfo {author}
  {\bibfnamefont {Y.}~\bibnamefont {Canvel}}, \bibinfo {author} {\bibfnamefont
  {M.}~\bibnamefont {Demarets}}, \bibinfo {author} {\bibfnamefont
  {D.}~\bibnamefont {Vangoidsenhoven}}, \bibinfo {author} {\bibfnamefont
  {Y.}~\bibnamefont {Hermans}}, \bibinfo {author} {\bibfnamefont {J.-G.}\
  \bibnamefont {Lai}}, \bibinfo {author} {\bibfnamefont {V.}~\bibnamefont
  {Rao}}, \bibinfo {author} {\bibfnamefont {M.}~\bibnamefont {Mongillo}},
  \bibinfo {author} {\bibfnamefont {D.}~\bibnamefont {Wan}}, \bibinfo {author}
  {\bibfnamefont {J.~D.}\ \bibnamefont {Boeck}}, \bibinfo {author}
  {\bibfnamefont {A.}~\bibnamefont {Potocnik}},\ and\ \bibinfo {author}
  {\bibfnamefont {K.~D.}\ \bibnamefont {Greve}},\ }\href@noop {} {\bibinfo
  {title} {High-coherence superconducting qubits made using industry-standard,
  advanced semiconductor manufacturing}} (\bibinfo {year} {2024}),\ \Eprint
  {https://arxiv.org/abs/2403.01312} {arXiv:2403.01312 [quant-ph]} \BibitemShut
  {NoStop}%
\bibitem [{\citenamefont {Krause}\ \emph {et~al.}(2024)\citenamefont {Krause},
  \citenamefont {Marchegiani}, \citenamefont {Janssen}, \citenamefont
  {Catelani}, \citenamefont {Ando},\ and\ \citenamefont {Dickel}}]{Krause2024}%
  \BibitemOpen
  \bibfield  {author} {\bibinfo {author} {\bibfnamefont {J.}~\bibnamefont
  {Krause}}, \bibinfo {author} {\bibfnamefont {G.}~\bibnamefont {Marchegiani}},
  \bibinfo {author} {\bibfnamefont {L.~M.}\ \bibnamefont {Janssen}}, \bibinfo
  {author} {\bibfnamefont {G.}~\bibnamefont {Catelani}}, \bibinfo {author}
  {\bibfnamefont {Y.}~\bibnamefont {Ando}},\ and\ \bibinfo {author}
  {\bibfnamefont {C.}~\bibnamefont {Dickel}},\ }\href@noop {} {\bibinfo {title}
  {Quasiparticle effects in magnetic-field-resilient 3d transmons}} (\bibinfo
  {year} {2024}),\ \Eprint {https://arxiv.org/abs/2403.03351} {arXiv:2403.03351
  [quant-ph]} \BibitemShut {NoStop}%
\bibitem [{\citenamefont {Houzet}\ \emph {et~al.}(2019)\citenamefont {Houzet},
  \citenamefont {Serniak}, \citenamefont {Catelani}, \citenamefont {Devoret},\
  and\ \citenamefont {Glazman}}]{Houzet2019}%
  \BibitemOpen
  \bibfield  {author} {\bibinfo {author} {\bibfnamefont {M.}~\bibnamefont
  {Houzet}}, \bibinfo {author} {\bibfnamefont {K.}~\bibnamefont {Serniak}},
  \bibinfo {author} {\bibfnamefont {G.}~\bibnamefont {Catelani}}, \bibinfo
  {author} {\bibfnamefont {M.~H.}\ \bibnamefont {Devoret}},\ and\ \bibinfo
  {author} {\bibfnamefont {L.~I.}\ \bibnamefont {Glazman}},\ }\href
  {https://doi.org/10.1103/PhysRevLett.123.107704} {\bibfield  {journal}
  {\bibinfo  {journal} {Phys. Rev. Lett.}\ }\textbf {\bibinfo {volume} {123}},\
  \bibinfo {pages} {107704} (\bibinfo {year} {2019})}\BibitemShut {NoStop}%
\bibitem [{\citenamefont {Saira}\ \emph {et~al.}(2012)\citenamefont {Saira},
  \citenamefont {Kemppinen}, \citenamefont {Maisi},\ and\ \citenamefont
  {Pekola}}]{Saira2012}%
  \BibitemOpen
  \bibfield  {author} {\bibinfo {author} {\bibfnamefont {O.-P.}\ \bibnamefont
  {Saira}}, \bibinfo {author} {\bibfnamefont {A.}~\bibnamefont {Kemppinen}},
  \bibinfo {author} {\bibfnamefont {V.~F.}\ \bibnamefont {Maisi}},\ and\
  \bibinfo {author} {\bibfnamefont {J.~P.}\ \bibnamefont {Pekola}},\ }\href
  {https://doi.org/10.1103/PhysRevB.85.012504} {\bibfield  {journal} {\bibinfo
  {journal} {Physical Review B}\ }\textbf {\bibinfo {volume} {85}},\ \bibinfo
  {pages} {012504} (\bibinfo {year} {2012})}\BibitemShut {NoStop}%
\bibitem [{\citenamefont {{van Woerkom}}\ \emph {et~al.}(2015)\citenamefont
  {{van Woerkom}}, \citenamefont {Geresdi},\ and\ \citenamefont
  {Kouwenhoven}}]{VanWoerkom2015}%
  \BibitemOpen
  \bibfield  {author} {\bibinfo {author} {\bibfnamefont {D.~J.}\ \bibnamefont
  {{van Woerkom}}}, \bibinfo {author} {\bibfnamefont {A.}~\bibnamefont
  {Geresdi}},\ and\ \bibinfo {author} {\bibfnamefont {L.~P.}\ \bibnamefont
  {Kouwenhoven}},\ }\href {https://www.nature.com/articles/nphys3342}
  {\bibfield  {journal} {\bibinfo  {journal} {Nature Physics}\ }\textbf
  {\bibinfo {volume} {11}},\ \bibinfo {pages} {547} (\bibinfo {year}
  {2015})}\BibitemShut {NoStop}%
\bibitem [{\citenamefont {McEwen}\ \emph {et~al.}(2024)\citenamefont {McEwen},
  \citenamefont {Miao}, \citenamefont {Atalaya}, \citenamefont {Bilmes},
  \citenamefont {Crook}, \citenamefont {Bovaird}, \citenamefont {Kreikebaum},
  \citenamefont {Zobrist}, \citenamefont {Jeffrey}, \citenamefont {Ying},
  \citenamefont {Bengtsson}, \citenamefont {Chang}, \citenamefont {Dunsworth},
  \citenamefont {Kelly}, \citenamefont {Zhang}, \citenamefont {Forati},
  \citenamefont {Acharya}, \citenamefont {Iveland}, \citenamefont {Liu},
  \citenamefont {Kim}, \citenamefont {Burkett}, \citenamefont {Megrant},
  \citenamefont {Chen}, \citenamefont {Neill}, \citenamefont {Sank},
  \citenamefont {Devoret},\ and\ \citenamefont {Opremcak}}]{McEwen2024}%
  \BibitemOpen
  \bibfield  {author} {\bibinfo {author} {\bibfnamefont {M.}~\bibnamefont
  {McEwen}}, \bibinfo {author} {\bibfnamefont {K.~C.}\ \bibnamefont {Miao}},
  \bibinfo {author} {\bibfnamefont {J.}~\bibnamefont {Atalaya}}, \bibinfo
  {author} {\bibfnamefont {A.}~\bibnamefont {Bilmes}}, \bibinfo {author}
  {\bibfnamefont {A.}~\bibnamefont {Crook}}, \bibinfo {author} {\bibfnamefont
  {J.}~\bibnamefont {Bovaird}}, \bibinfo {author} {\bibfnamefont {J.~M.}\
  \bibnamefont {Kreikebaum}}, \bibinfo {author} {\bibfnamefont
  {N.}~\bibnamefont {Zobrist}}, \bibinfo {author} {\bibfnamefont
  {E.}~\bibnamefont {Jeffrey}}, \bibinfo {author} {\bibfnamefont
  {B.}~\bibnamefont {Ying}}, \bibinfo {author} {\bibfnamefont {A.}~\bibnamefont
  {Bengtsson}}, \bibinfo {author} {\bibfnamefont {H.-S.}\ \bibnamefont
  {Chang}}, \bibinfo {author} {\bibfnamefont {A.}~\bibnamefont {Dunsworth}},
  \bibinfo {author} {\bibfnamefont {J.}~\bibnamefont {Kelly}}, \bibinfo
  {author} {\bibfnamefont {Y.}~\bibnamefont {Zhang}}, \bibinfo {author}
  {\bibfnamefont {E.}~\bibnamefont {Forati}}, \bibinfo {author} {\bibfnamefont
  {R.}~\bibnamefont {Acharya}}, \bibinfo {author} {\bibfnamefont
  {J.}~\bibnamefont {Iveland}}, \bibinfo {author} {\bibfnamefont
  {W.}~\bibnamefont {Liu}}, \bibinfo {author} {\bibfnamefont {S.}~\bibnamefont
  {Kim}}, \bibinfo {author} {\bibfnamefont {B.}~\bibnamefont {Burkett}},
  \bibinfo {author} {\bibfnamefont {A.}~\bibnamefont {Megrant}}, \bibinfo
  {author} {\bibfnamefont {Y.}~\bibnamefont {Chen}}, \bibinfo {author}
  {\bibfnamefont {C.}~\bibnamefont {Neill}}, \bibinfo {author} {\bibfnamefont
  {D.}~\bibnamefont {Sank}}, \bibinfo {author} {\bibfnamefont {M.}~\bibnamefont
  {Devoret}},\ and\ \bibinfo {author} {\bibfnamefont {A.}~\bibnamefont
  {Opremcak}},\ }\href@noop {} {\bibinfo {title} {Resisting high-energy impact
  events through gap engineering in superconducting qubit arrays}} (\bibinfo
  {year} {2024}),\ \Eprint {https://arxiv.org/abs/2402.15644} {arxiv:2402.15644
  [quant-ph]} \BibitemShut {NoStop}%
\bibitem [{\citenamefont {202Q-lab}(2023)}]{202Qlab2023}%
  \BibitemOpen
  \bibfield  {author} {\bibinfo {author} {\bibnamefont {202Q-lab}},\ }\href
  {https://github.com/202Q-lab/OQTO} {\bibinfo {title} {{OQTO: 8-port Cryogenic
  Sample Holder}}} (\bibinfo {year} {2023}),\ \bibinfo {note} {gitHub
  repository}\BibitemShut {NoStop}%
\bibitem [{\citenamefont {Walter}\ \emph {et~al.}(2017)\citenamefont {Walter},
  \citenamefont {Kurpiers}, \citenamefont {Gasparinetti}, \citenamefont
  {Magnard}, \citenamefont {Poto{\v c}nik}, \citenamefont {Salath{\'e}},
  \citenamefont {Pechal}, \citenamefont {Mondal}, \citenamefont {Oppliger},
  \citenamefont {Eichler},\ and\ \citenamefont {Wallraff}}]{Walter2017}%
  \BibitemOpen
  \bibfield  {author} {\bibinfo {author} {\bibfnamefont {T.}~\bibnamefont
  {Walter}}, \bibinfo {author} {\bibfnamefont {P.}~\bibnamefont {Kurpiers}},
  \bibinfo {author} {\bibfnamefont {S.}~\bibnamefont {Gasparinetti}}, \bibinfo
  {author} {\bibfnamefont {P.}~\bibnamefont {Magnard}}, \bibinfo {author}
  {\bibfnamefont {A.}~\bibnamefont {Poto{\v c}nik}}, \bibinfo {author}
  {\bibfnamefont {Y.}~\bibnamefont {Salath{\'e}}}, \bibinfo {author}
  {\bibfnamefont {M.}~\bibnamefont {Pechal}}, \bibinfo {author} {\bibfnamefont
  {M.}~\bibnamefont {Mondal}}, \bibinfo {author} {\bibfnamefont
  {M.}~\bibnamefont {Oppliger}}, \bibinfo {author} {\bibfnamefont
  {C.}~\bibnamefont {Eichler}},\ and\ \bibinfo {author} {\bibfnamefont
  {A.}~\bibnamefont {Wallraff}},\ }\href
  {https://doi.org/10.1103/PhysRevApplied.7.054020} {\bibfield  {journal}
  {\bibinfo  {journal} {Phys. Rev. Applied}\ }\textbf {\bibinfo {volume} {7}},\
  \bibinfo {pages} {054020} (\bibinfo {year} {2017})}\BibitemShut {NoStop}%
\bibitem [{\citenamefont {Lecocq}\ \emph {et~al.}(2020)\citenamefont {Lecocq},
  \citenamefont {Ranzani}, \citenamefont {Peterson}, \citenamefont {Cicak},
  \citenamefont {Metelmann}, \citenamefont {Kotler}, \citenamefont {Simmonds},
  \citenamefont {Teufel},\ and\ \citenamefont {Aumentado}}]{Lecocq2020}%
  \BibitemOpen
  \bibfield  {author} {\bibinfo {author} {\bibfnamefont {F.}~\bibnamefont
  {Lecocq}}, \bibinfo {author} {\bibfnamefont {L.}~\bibnamefont {Ranzani}},
  \bibinfo {author} {\bibfnamefont {G.}~\bibnamefont {Peterson}}, \bibinfo
  {author} {\bibfnamefont {K.}~\bibnamefont {Cicak}}, \bibinfo {author}
  {\bibfnamefont {A.}~\bibnamefont {Metelmann}}, \bibinfo {author}
  {\bibfnamefont {S.}~\bibnamefont {Kotler}}, \bibinfo {author} {\bibfnamefont
  {R.}~\bibnamefont {Simmonds}}, \bibinfo {author} {\bibfnamefont
  {J.}~\bibnamefont {Teufel}},\ and\ \bibinfo {author} {\bibfnamefont
  {J.}~\bibnamefont {Aumentado}},\ }\href
  {https://doi.org/10.1103/PhysRevApplied.13.044005} {\bibfield  {journal}
  {\bibinfo  {journal} {Phys. Rev. Appl.}\ }\textbf {\bibinfo {volume} {13}},\
  \bibinfo {pages} {044005} (\bibinfo {year} {2020})}\BibitemShut {NoStop}%
\bibitem [{\citenamefont {Ranadive}\ \emph {et~al.}(2022)\citenamefont
  {Ranadive}, \citenamefont {Esposito}, \citenamefont {Planat}, \citenamefont
  {Bonet}, \citenamefont {Naud}, \citenamefont {Buisson}, \citenamefont
  {Guichard},\ and\ \citenamefont {Roch}}]{Ranadive2022}%
  \BibitemOpen
  \bibfield  {author} {\bibinfo {author} {\bibfnamefont {A.}~\bibnamefont
  {Ranadive}}, \bibinfo {author} {\bibfnamefont {M.}~\bibnamefont {Esposito}},
  \bibinfo {author} {\bibfnamefont {L.}~\bibnamefont {Planat}}, \bibinfo
  {author} {\bibfnamefont {E.}~\bibnamefont {Bonet}}, \bibinfo {author}
  {\bibfnamefont {C.}~\bibnamefont {Naud}}, \bibinfo {author} {\bibfnamefont
  {O.}~\bibnamefont {Buisson}}, \bibinfo {author} {\bibfnamefont
  {W.}~\bibnamefont {Guichard}},\ and\ \bibinfo {author} {\bibfnamefont
  {N.}~\bibnamefont {Roch}},\ }\href
  {https://doi.org/10.1038/s41467-022-29375-5} {\bibfield  {journal} {\bibinfo
  {journal} {Nature Communications}\ }\textbf {\bibinfo {volume} {13}},\
  \bibinfo {pages} {1737} (\bibinfo {year} {2022})},\ \Eprint
  {https://arxiv.org/abs/2101.05815} {arxiv:2101.05815} \BibitemShut {NoStop}%
\bibitem [{\citenamefont {Harrington}\ \emph {et~al.}(2024)\citenamefont
  {Harrington}, \citenamefont {Li}, \citenamefont {Hays}, \citenamefont {Van
  De~Pontseele}, \citenamefont {Mayer}, \citenamefont {Pinckney}, \citenamefont
  {Contipelli}, \citenamefont {Gingras}, \citenamefont {Niedzielski},
  \citenamefont {Stickler}, \citenamefont {Yoder}, \citenamefont {Schwartz},
  \citenamefont {Grover}, \citenamefont {Serniak}, \citenamefont {Oliver},\
  and\ \citenamefont {Formaggio}}]{Harrington2024}%
  \BibitemOpen
  \bibfield  {author} {\bibinfo {author} {\bibfnamefont {P.~M.}\ \bibnamefont
  {Harrington}}, \bibinfo {author} {\bibfnamefont {M.}~\bibnamefont {Li}},
  \bibinfo {author} {\bibfnamefont {M.}~\bibnamefont {Hays}}, \bibinfo {author}
  {\bibfnamefont {W.}~\bibnamefont {Van De~Pontseele}}, \bibinfo {author}
  {\bibfnamefont {D.}~\bibnamefont {Mayer}}, \bibinfo {author} {\bibfnamefont
  {H.~D.}\ \bibnamefont {Pinckney}}, \bibinfo {author} {\bibfnamefont
  {F.}~\bibnamefont {Contipelli}}, \bibinfo {author} {\bibfnamefont
  {M.}~\bibnamefont {Gingras}}, \bibinfo {author} {\bibfnamefont {B.~M.}\
  \bibnamefont {Niedzielski}}, \bibinfo {author} {\bibfnamefont
  {H.}~\bibnamefont {Stickler}}, \bibinfo {author} {\bibfnamefont {J.~L.}\
  \bibnamefont {Yoder}}, \bibinfo {author} {\bibfnamefont {M.~E.}\ \bibnamefont
  {Schwartz}}, \bibinfo {author} {\bibfnamefont {J.~A.}\ \bibnamefont
  {Grover}}, \bibinfo {author} {\bibfnamefont {K.}~\bibnamefont {Serniak}},
  \bibinfo {author} {\bibfnamefont {W.~D.}\ \bibnamefont {Oliver}},\ and\
  \bibinfo {author} {\bibfnamefont {J.~A.}\ \bibnamefont {Formaggio}},\
  }\href@noop {} {\bibinfo {title} {Synchronous {{Detection}} of {{Cosmic
  Rays}} and {{Correlated Errors}} in {{Superconducting Qubit Arrays}}}}
  (\bibinfo {year} {2024}),\ \Eprint {https://arxiv.org/abs/2402.03208}
  {arxiv:2402.03208 [hep-ex, physics:nucl-ex, physics:physics,
  physics:quant-ph]} \BibitemShut {NoStop}%
\bibitem [{\citenamefont {Burnett}\ \emph {et~al.}(2019)\citenamefont
  {Burnett}, \citenamefont {Bengtsson}, \citenamefont {Scigliuzzo},
  \citenamefont {Niepce}, \citenamefont {Kudra}, \citenamefont {Delsing},\ and\
  \citenamefont {Bylander}}]{Burnett2019}%
  \BibitemOpen
  \bibfield  {author} {\bibinfo {author} {\bibfnamefont {J.~J.}\ \bibnamefont
  {Burnett}}, \bibinfo {author} {\bibfnamefont {A.}~\bibnamefont {Bengtsson}},
  \bibinfo {author} {\bibfnamefont {M.}~\bibnamefont {Scigliuzzo}}, \bibinfo
  {author} {\bibfnamefont {D.}~\bibnamefont {Niepce}}, \bibinfo {author}
  {\bibfnamefont {M.}~\bibnamefont {Kudra}}, \bibinfo {author} {\bibfnamefont
  {P.}~\bibnamefont {Delsing}},\ and\ \bibinfo {author} {\bibfnamefont
  {J.}~\bibnamefont {Bylander}},\ }\href
  {https://doi.org/10.1038/s41534-019-0168-5} {\bibfield  {journal} {\bibinfo
  {journal} {npj Quantum Information}\ }\textbf {\bibinfo {volume} {5}},\
  \bibinfo {pages} {54} (\bibinfo {year} {2019})}\BibitemShut {NoStop}%
\bibitem [{\citenamefont {Acharya}\ \emph {et~al.}(2023)\citenamefont
  {Acharya}, \citenamefont {Brebels}, \citenamefont {Grill}, \citenamefont
  {Verjauw}, \citenamefont {Ivanov}, \citenamefont {Lozano}, \citenamefont
  {Wan}, \citenamefont {Van~Damme}, \citenamefont {Vadiraj}, \citenamefont
  {Mongillo}, \citenamefont {Govoreanu}, \citenamefont {Craninckx},
  \citenamefont {Radu}, \citenamefont {De~Greve}, \citenamefont {Gielen},
  \citenamefont {Catthoor},\ and\ \citenamefont {Potočnik}}]{Acharya2023}%
  \BibitemOpen
  \bibfield  {author} {\bibinfo {author} {\bibfnamefont {R.}~\bibnamefont
  {Acharya}}, \bibinfo {author} {\bibfnamefont {S.}~\bibnamefont {Brebels}},
  \bibinfo {author} {\bibfnamefont {A.}~\bibnamefont {Grill}}, \bibinfo
  {author} {\bibfnamefont {J.}~\bibnamefont {Verjauw}}, \bibinfo {author}
  {\bibfnamefont {T.}~\bibnamefont {Ivanov}}, \bibinfo {author} {\bibfnamefont
  {D.~P.}\ \bibnamefont {Lozano}}, \bibinfo {author} {\bibfnamefont
  {D.}~\bibnamefont {Wan}}, \bibinfo {author} {\bibfnamefont {J.}~\bibnamefont
  {Van~Damme}}, \bibinfo {author} {\bibfnamefont {A.~M.}\ \bibnamefont
  {Vadiraj}}, \bibinfo {author} {\bibfnamefont {M.}~\bibnamefont {Mongillo}},
  \bibinfo {author} {\bibfnamefont {B.}~\bibnamefont {Govoreanu}}, \bibinfo
  {author} {\bibfnamefont {J.}~\bibnamefont {Craninckx}}, \bibinfo {author}
  {\bibfnamefont {I.~P.}\ \bibnamefont {Radu}}, \bibinfo {author}
  {\bibfnamefont {K.}~\bibnamefont {De~Greve}}, \bibinfo {author}
  {\bibfnamefont {G.}~\bibnamefont {Gielen}}, \bibinfo {author} {\bibfnamefont
  {F.}~\bibnamefont {Catthoor}},\ and\ \bibinfo {author} {\bibfnamefont
  {A.}~\bibnamefont {Potočnik}},\ }\href
  {https://doi.org/10.1038/s41928-023-01033-8} {\bibfield  {journal} {\bibinfo
  {journal} {Nature Electronics}\ }\textbf {\bibinfo {volume} {6}},\ \bibinfo
  {pages} {900} (\bibinfo {year} {2023})}\BibitemShut {NoStop}%
\bibitem [{\citenamefont {Pechal}\ \emph {et~al.}(2016)\citenamefont {Pechal},
  \citenamefont {Besse}, \citenamefont {Mondal}, \citenamefont {Oppliger},
  \citenamefont {Gasparinetti},\ and\ \citenamefont {Wallraff}}]{Pechal2016}%
  \BibitemOpen
  \bibfield  {author} {\bibinfo {author} {\bibfnamefont {M.}~\bibnamefont
  {Pechal}}, \bibinfo {author} {\bibfnamefont {J.-C.}\ \bibnamefont {Besse}},
  \bibinfo {author} {\bibfnamefont {M.}~\bibnamefont {Mondal}}, \bibinfo
  {author} {\bibfnamefont {M.}~\bibnamefont {Oppliger}}, \bibinfo {author}
  {\bibfnamefont {S.}~\bibnamefont {Gasparinetti}},\ and\ \bibinfo {author}
  {\bibfnamefont {A.}~\bibnamefont {Wallraff}},\ }\href
  {https://doi.org/10.1103/PhysRevApplied.6.024009} {\bibfield  {journal}
  {\bibinfo  {journal} {Phys. Rev. Appl.}\ }\textbf {\bibinfo {volume} {6}},\
  \bibinfo {pages} {024009} (\bibinfo {year} {2016})}\BibitemShut {NoStop}%
\bibitem [{\citenamefont {Ritter}\ \emph {et~al.}(2021)\citenamefont {Ritter},
  \citenamefont {Fuhrer}, \citenamefont {Haxell}, \citenamefont {Hart},
  \citenamefont {Gumann}, \citenamefont {Riel},\ and\ \citenamefont
  {Nichele}}]{Ritter2021}%
  \BibitemOpen
  \bibfield  {author} {\bibinfo {author} {\bibfnamefont {M.~F.}\ \bibnamefont
  {Ritter}}, \bibinfo {author} {\bibfnamefont {A.}~\bibnamefont {Fuhrer}},
  \bibinfo {author} {\bibfnamefont {D.~Z.}\ \bibnamefont {Haxell}}, \bibinfo
  {author} {\bibfnamefont {S.}~\bibnamefont {Hart}}, \bibinfo {author}
  {\bibfnamefont {P.}~\bibnamefont {Gumann}}, \bibinfo {author} {\bibfnamefont
  {H.}~\bibnamefont {Riel}},\ and\ \bibinfo {author} {\bibfnamefont
  {F.}~\bibnamefont {Nichele}},\ }\href
  {https://doi.org/10.1038/s41467-021-21231-2} {\bibfield  {journal} {\bibinfo
  {journal} {Nature Communications}\ }\textbf {\bibinfo {volume} {12}},\
  \bibinfo {pages} {1266} (\bibinfo {year} {2021})}\BibitemShut {NoStop}%
\bibitem [{\citenamefont {Rehammar}\ and\ \citenamefont
  {Gasparinetti}(2023)}]{Rehammar2023}%
  \BibitemOpen
  \bibfield  {author} {\bibinfo {author} {\bibfnamefont {R.}~\bibnamefont
  {Rehammar}}\ and\ \bibinfo {author} {\bibfnamefont {S.}~\bibnamefont
  {Gasparinetti}},\ }\href {https://doi.org/10.1109/TMTT.2023.3238543}
  {\bibfield  {journal} {\bibinfo  {journal} {IEEE Transactions on Microwave
  Theory and Techniques}\ }\textbf {\bibinfo {volume} {71}},\ \bibinfo {pages}
  {3075} (\bibinfo {year} {2023})}\BibitemShut {NoStop}%
\bibitem [{\citenamefont {Tholén}\ \emph {et~al.}(2022)\citenamefont
  {Tholén}, \citenamefont {Borgani}, \citenamefont {Di~Carlo}, \citenamefont
  {Bengtsson}, \citenamefont {Križan}, \citenamefont {Kudra}, \citenamefont
  {Tancredi}, \citenamefont {Bylander}, \citenamefont {Delsing}, \citenamefont
  {Gasparinetti},\ and\ \citenamefont {Haviland}}]{Tholen2022}%
  \BibitemOpen
  \bibfield  {author} {\bibinfo {author} {\bibfnamefont {M.~O.}\ \bibnamefont
  {Tholén}}, \bibinfo {author} {\bibfnamefont {R.}~\bibnamefont {Borgani}},
  \bibinfo {author} {\bibfnamefont {G.~R.}\ \bibnamefont {Di~Carlo}}, \bibinfo
  {author} {\bibfnamefont {A.}~\bibnamefont {Bengtsson}}, \bibinfo {author}
  {\bibfnamefont {C.}~\bibnamefont {Križan}}, \bibinfo {author} {\bibfnamefont
  {M.}~\bibnamefont {Kudra}}, \bibinfo {author} {\bibfnamefont
  {G.}~\bibnamefont {Tancredi}}, \bibinfo {author} {\bibfnamefont
  {J.}~\bibnamefont {Bylander}}, \bibinfo {author} {\bibfnamefont
  {P.}~\bibnamefont {Delsing}}, \bibinfo {author} {\bibfnamefont
  {S.}~\bibnamefont {Gasparinetti}},\ and\ \bibinfo {author} {\bibfnamefont
  {D.~B.}\ \bibnamefont {Haviland}},\ }\href
  {https://doi.org/10.1063/5.0101398} {\bibfield  {journal} {\bibinfo
  {journal} {Review of Scientific Instruments}\ }\textbf {\bibinfo {volume}
  {93}},\ \bibinfo {pages} {104711} (\bibinfo {year} {2022})}\BibitemShut
  {NoStop}%
\bibitem [{\citenamefont {Kurter}\ \emph {et~al.}(2022)\citenamefont {Kurter},
  \citenamefont {Murray}, \citenamefont {Gordon}, \citenamefont {Wymore},
  \citenamefont {Sandberg}, \citenamefont {Shelby}, \citenamefont {Eddins},
  \citenamefont {Adiga}, \citenamefont {Finck}, \citenamefont {Rivera},
  \citenamefont {Stabile}, \citenamefont {Trimm}, \citenamefont {Wacaser},
  \citenamefont {Balakrishnan}, \citenamefont {Pyzyna}, \citenamefont
  {Sleight}, \citenamefont {Steffen},\ and\ \citenamefont
  {Rodbell}}]{Kurter2022}%
  \BibitemOpen
  \bibfield  {author} {\bibinfo {author} {\bibfnamefont {C.}~\bibnamefont
  {Kurter}}, \bibinfo {author} {\bibfnamefont {C.~E.}\ \bibnamefont {Murray}},
  \bibinfo {author} {\bibfnamefont {R.~T.}\ \bibnamefont {Gordon}}, \bibinfo
  {author} {\bibfnamefont {B.~B.}\ \bibnamefont {Wymore}}, \bibinfo {author}
  {\bibfnamefont {M.}~\bibnamefont {Sandberg}}, \bibinfo {author}
  {\bibfnamefont {R.~M.}\ \bibnamefont {Shelby}}, \bibinfo {author}
  {\bibfnamefont {A.}~\bibnamefont {Eddins}}, \bibinfo {author} {\bibfnamefont
  {V.~P.}\ \bibnamefont {Adiga}}, \bibinfo {author} {\bibfnamefont {A.~D.~K.}\
  \bibnamefont {Finck}}, \bibinfo {author} {\bibfnamefont {E.}~\bibnamefont
  {Rivera}}, \bibinfo {author} {\bibfnamefont {A.~A.}\ \bibnamefont {Stabile}},
  \bibinfo {author} {\bibfnamefont {B.}~\bibnamefont {Trimm}}, \bibinfo
  {author} {\bibfnamefont {B.}~\bibnamefont {Wacaser}}, \bibinfo {author}
  {\bibfnamefont {K.}~\bibnamefont {Balakrishnan}}, \bibinfo {author}
  {\bibfnamefont {A.}~\bibnamefont {Pyzyna}}, \bibinfo {author} {\bibfnamefont
  {J.}~\bibnamefont {Sleight}}, \bibinfo {author} {\bibfnamefont
  {M.}~\bibnamefont {Steffen}},\ and\ \bibinfo {author} {\bibfnamefont
  {K.}~\bibnamefont {Rodbell}},\ }\href
  {https://doi.org/10.1038/s41534-022-00542-2} {\bibfield  {journal} {\bibinfo
  {journal} {npj Quantum Information}\ }\textbf {\bibinfo {volume} {8}},\
  \bibinfo {pages} {1} (\bibinfo {year} {2022})}\BibitemShut {NoStop}%
\bibitem [{\citenamefont {Marchegiani}\ \emph {et~al.}(2022)\citenamefont
  {Marchegiani}, \citenamefont {Amico},\ and\ \citenamefont
  {Catelani}}]{Marchegiani2022}%
  \BibitemOpen
  \bibfield  {author} {\bibinfo {author} {\bibfnamefont {G.}~\bibnamefont
  {Marchegiani}}, \bibinfo {author} {\bibfnamefont {L.}~\bibnamefont {Amico}},\
  and\ \bibinfo {author} {\bibfnamefont {G.}~\bibnamefont {Catelani}},\ }\href
  {https://doi.org/10.1103/PRXQuantum.3.040338} {\bibfield  {journal} {\bibinfo
   {journal} {PRX Quantum}\ }\textbf {\bibinfo {volume} {3}},\ \bibinfo {pages}
  {040338} (\bibinfo {year} {2022})}\BibitemShut {NoStop}%
\end{thebibliography}

%

\end{document}